\documentclass[a4paper,11pt]{article}
\usepackage{jheppub}

\newcommand{\be}{\begin{equation}}
\newcommand{\ee}{\end{equation}}
\newcommand{\ba}{\begin{eqnarray}}
\newcommand{\ea}{\end{eqnarray}}
\newcommand{\la}{\label}
\newcommand\Tr{\mbox{Tr}\, }
\def\sumint{\hbox{$\sum$}\!\!\!\!\!\!\int}

\title{\boldmath Two-loop perturbative corrections to the constrained effective potential
in thermal QCD}

\author[a,b]{Yun Guo}
\author[c,a]{Qianqian Du}

\affiliation[a]{Physics Department, Guangxi Normal University, Guilin   541004, P.R.C. }
\affiliation[b]{Guangxi Key Laboratory of Nuclear Physics and Technology, Guilin   541004, P.R.C. }
\affiliation[c]{Institute of Particle Physics and Key Laboratory of Quark and Lepton Physics (MOS) ,  \\ Central China Normal University,  Wuhan,  430079, P.R.C.} %

\emailAdd{yunguo@mailbox.gxnu.edu.cn}
\emailAdd{duqianqianstudent@mails.ccnu.edu.cn}

\abstract{
In this paper, we compute the constrained QCD effective potential up to two-loop order
with finite quark mass and chemical potential. We present the explicit calculations by
using the double line notation and analytical expressions for massless quarks are obtained
in terms of the Bernoulli polynomials or Polyakov loops. Our results explicitly show that
the constrained QCD effective potential is independent on the gauge fixing parameter.
In addition, as compared to the massless case, the constrained QCD effective potential
with massive quarks develops a completely new term which is only absent when the
background field vanishes. Furthermore, we discuss the relation between the one- and
two- loop constrained effective potential. The surprisingly simple proportionality that
exists in the pure gauge theories, however, is in general no longer true when fermions are
taken into account. On the other hand, for high baryon density $\mu_B$ and low temperature
$T$, in the massless limit, we do also find a similar proportionality between the one- and
two-loop fermionic contributions in the constrained effective potential up to ${\cal O}(T/\mu_B)$.
}

\keywords{
 Constrained effective potential, Polyakov loop, Thermal field theory
}

\begin{document}
\maketitle
\flushbottom

\section{Introduction}\la{intro}

Many exciting experimental phenomena observed at RHIC and the LHC have confirmed the formation of the Quark-Gluon Plasma(QGP) in the extremely hot and dense condition\cite{Gyulassy:2004zy,Shuryak:2008eq,Adcox:2004mh,Adams:2005dq}. Understanding the phase transition from the normal hadronic matter to QGP and studying the equation of state(EOS) of the deconfined matter are of great interest in heavy ion physics and also important for astrophysics and cosmology. Quantum ChromoDynamics(QCD) is the theory that describes the strong interactions. At the asymptotic high temperatures, one could perturbatively calculate the EOS from first principles\cite{Gross:1973id,Braaten:1989mz}. Near the phase transition temperature, however, non-perturbative physics becomes very important and lattice QCD simulation is a reliable theoretical tool that can provide useful information on the thermodynamics of QGP, see Refs.~\cite{Boyd:1996bx,Petreczky:2012rq,Borsanyi:2012ve,Panero:2009tv,Datta:2010sq,Lucini:2003zr,Bazavov:2009zn,Borsanyi:2010cj} for examples. Based on the lattice simulations, constructing phenomenological models to study the phase transition has made a lot of progress over last decades.

Recently, matrix models for deconfinement have been proposed which, with a relatively simple form, reproduce the lattice results in the phase transition region very well\cite{Meisinger:2001cq,Dumitru:2010mj,Dumitru:2012fw,Guo:2014zra}. They are also generalized to the QCD case by including quarks\cite{Pisarski:2016ixt}. The basic structures of the matrix models depend on the effective potential. Namely, they can be obtained by a high temperature expansion of the effective potential with a modified dispersion relation of gauge bosons. 
At present, all of these models use the one-loop effective potential as the ideal contributions. Therefore, studying the perturbative corrections to the effective potential can not only improve the high temperature behavior of the thermodynamics but also give rise to possible modifications on the non-ideal terms in matrix models.

The constrained effective potential is simply the (logarithm of) partition function with the traditional action in the path integral replaced by its constrained version. It results from the requirement that path integral over the gauge fields is performed while preserving the value of the Polyakov loop at some fixed value\cite{ORaifeartaigh:1986axd,Bhattacharya:1992qb,KorthalsAltes:1993ca}. We define the constrained effective potential $\Gamma$ by the following equation \footnote{In the following of this paper, in most cases, $\Gamma$ is called effective potential for short which of course refers to the constrained version as defined in Eq.~(\ref{11}). The fermionic fields can be added straightforwardly. To keep the notations compact, we don't include them in this section.}
\be\la{11}
\exp (- V \beta \Gamma({\ell}_k)) = \int
D A_\mu \prod^{N-1}_{k=1}\delta({\ell}_k-\bar {\ell}_k)\exp \bigg(-\frac{S(A)}{g^2}\bigg)\, ,
\ee
where $V$ is the volume of the system under consideration, $\beta\equiv 1/T$ is the inverse temperature, $g$ is the strong coupling constant and $N$ is the number of colors. The action for the gauge field is given by $S(A)=\int_0^{\beta} d \tau \int d^3\vec x \,\Tr (G^2_{\mu \nu}/2)$ where $G_{\mu\nu}=\partial _\mu A_\nu-\partial _\nu A_\mu-i[A_\mu,A_\nu]$. The spatial average of the trace of ($k$-th powers of the) Polyakov loop ${\bf L}(\vec x)$ is denoted as $\bar {\ell}_k$, which is given by
\be
\bar {\ell}_k\equiv \frac{1}{N} \overline{\Tr{\bf L}^k}={1\over N}\frac{\int_V d^3 \vec x \, \Tr {\bf L}^k (\vec x)}{V}\, ,
\ee
with
\be
{\bf L}(\vec x)={\cal P}\,
\exp\bigg( i\int_0^{\beta} d \tau A_0(\vec x,\tau) \bigg) \,,
\ee
where ${\cal P}$ is used to denote the path ordering. The delta functions in the above path integral give rise to a modification on the action $S(A)$ by a Fourier transform and this leads to the constrained version which reads
\be
S_{{\rm con}}(A,\epsilon) = i\sum^{N-1}_{k=1}{\epsilon_k}
\left({\ell}_k-\bar {\ell}_k\right) + \frac{S(A)}{g^2}\,.
\ee
Using the constrained action, $N-1$ integrals over the extra fields $\epsilon_k$ are introduced.

In order to compute the effective potential in a perturbative way, we need to expand the gauge fields and the introduced $\epsilon$ fields around some fixed classical values as
\be
A_\mu=A^{\rm cl}_\mu + g B_\mu\, , ~\mbox{and}~ \epsilon=\epsilon_c+g\epsilon_q\,,
\ee
where the corresponding quantum fluctuations are denoted by $B_\mu$ and $\epsilon_q$. To define the functional integral, we must gauge-fix by using the Faddeev-Popov procedure. Then one needs to add the gauge fixing and ghost contributions into the constrained effective action $S_{{\rm con}}$. The gauge fixing term is given by
\be\la{gf}
S_{gf}={1\over {\xi}}\int d^3\vec x ~ d\tau\, \Tr(D_{\mu}^{{\rm cl}}B_\mu)^2\,,
\ee
with $D_\mu^{{\rm cl}}=\partial_\mu-i[A^{\rm cl}_\mu\,,\cdots]$ being the classical covariant derivative in the adjoint representation. Eq.~(\ref{gf}) corresponds to the general covariant background gauge with gauge parameter $\xi$. In addition, the ghost contribution to the constrained action is
\be
S_{gh}= - 2 \int d^3\vec x ~ d\tau\, \Tr(\bar{\eta} D_{\mu}^{{\rm cl}}D_{\mu}\eta)\, ,
\ee
where $D_\mu=\partial_\mu-i[A_{\mu}\,,\cdots]$ is the covariant derivative in the adjoint representation.

Expand the constrained action in terms of the quantum fluctuations $B_\mu$ and $\epsilon_q$, the linear terms in the fluctuations are required to vanish which gives the saddle point equations. In our calculation, we consider a constant (classical) background field as $A^{\rm cl}_\mu=C \delta_{\mu 0}$. One can easily check that the corresponding saddle point equations are satisfied when $\epsilon_c$ is simply chosen to be zero. In addition, we can also get $\ell_k=\frac{1}{N}\Tr e^{i k \beta C}$. The terms quadratic in the fluctuations and the ghost fields correspond to the one-loop effective potential.
Considering the interactions, we can expand the constrained action to order $g^2$ which gives the two-loop result. Except for the usual free energy contribution $F$, the interaction terms in the expansion also involve the extra fields $\epsilon_k$ which give the insertion contribution at two and more loops. It corresponds to the radiative corrections inserted into the renormalized Polyakov loop and is denoted by $U$. According to the above discussion, we can formally express the effective potential as $\Gamma=F+U$.

In this paper, we perform a perturbative calculation for the QCD effective potential up to two-loop order by taking into account finite quark mass $m$ and chemical potential $\mu$. 
In general, the free-energy can be expressed as the momentum integrals where the integrand is given in terms of the parton distribution functions. At $A_0^{{\rm cl}}=0$, we simply have the Fermi-Dirac and Bose-Einstein distribution functions. For a non-zero $A_0^{{\rm cl}}$, the free energy becomes gauge-dependent. However, as shown by previous studies for pure gauge theories\cite{KorthalsAltes:1993ca,Dumitru:2013xna}, in Feynman gauge with $\xi=1$, the one- and two-loop free energy in a background field have the same structure as those computed at vanishing $A_0^{{\rm cl}}$ and the only change is a modification on the distribution functions which now depend on the background fields.
We are interested in the generalization of this conclusion to the fermionic contributions. In fact, the gluon self-energy in a constant $A_0^{{\rm cl}}$ shows some unexpected behaviors, namely it contains a new structure as compared to that at $A_0^{{\rm cl}}=0$\cite{Hidaka:2009hs}. Therefore, it is also possible to see the appearance of some new contributions in the effective potential when quark propagators set in.

In the large volume limit, the (constrained) effective potential defined in Eq.~(\ref{11}) is equivalent to the traditional definition of the effective potential where the source is coupled to the Polyakov loop\cite{Dumitru:2013xna}. Therefore, the effective potential is gauge invariant by construction. The gauge invariance of the gluonic effective potential at order $g^2$ was explicitly shown in Refs.~\cite{Belyaev:1991gh,KorthalsAltes:1993ca}. In this work, we will show the same result for full QCD effective potential with finite quark mass and chemical potential. This can be achieved with the explicit expressions for the free energy $F$ and the insertion contribution $U$ in general covariant background gauge since they directly enable us to see how the $\xi$-dependent terms in $F$ is cancelled by those terms in $U$.

Another motivation of this work is to study the relation between the one- and two-loop effective potential. For pure gauge theories, two-loop correction is proportional to the one-loop result, independent on the eigenvalues of the Polyakov loop\cite{Dumitru:2013xna,Guo:2014zra}. Therefore, the former takes a very simple form including only the periodic Bernoulli polynomial $B_4(x)$. A straightforward question is about the validity of the extension to fermionic sector in the effective potential. Even if such a simple relation doesn't hold after including the quark contributions, it is still interesting to look for some possible way to simplify the two-loop results for fermions.

The rest of the paper is organized as the following. In section \ref{review}, we review some basics of the double line basis and summarize the corresponding Feynman rules in a background field. 
In section \ref{fe}, for completeness, we first discuss the known one-loop free energy by introducing some new calculational techniques which demonstrate the corrections due to non-zero $A_0^{{\rm cl}}$ in a more clear way. Then we present the details for the calculation of the two-loop free energies in covariant gauge with an arbitrary gauge parameter $\xi$. Explicit results are given for each individual diagram and analytical expressions for massless quarks are obtained in terms of the periodic Bernoulli polynomials or Polyakov loops. In section \ref{in}, we compute the insertion contribution that arises due to the interactions involving the quantum fluctuations of the $\epsilon$ fields and explicitly show how the gauge dependence in the QCD free-energy is cancelled by that in the insertion contribution. In section \ref{rela}, we analyze the relation between the one- and two-loop result for fermionic effective potential and prove some simple proportionalities which are similar as those found in the pure gauge theories. A short summary is given in section \ref{summary}. In addition, the definition of the periodic Bernoulli polynomials as well as some important sum-integrals are discussed in the appendices.

\section{Double line basis and the corresponding Feynman rules}\la{review}

As compared to the usual Cartan basis, the double line basis is believed to be more efficient when compute in the presence of a constant $A^{{\rm cl}}_0$\cite{Cvitanovic:1976am}. For example, it has been used to compute the free-energies for $SU(\infty)$ gauge theory on a small sphere up to three-loop order\cite{Aharony:2005bq} and the quark/gluon self-energies for $SU(N)$ gauge theory at leading order\cite{Hidaka:2009hs}. In this section, we follow Refs.~\cite{Hidaka:2009hs,Cvitanovic:1976am} and give a very brief review on the double line basis and list the corresponding Feynman rules.

The generators of the fundamental representation are given by the projection operators,
\be
\la{21}
(t^{ab})_{cd} = \frac{1}{\sqrt{2}}{\cal P}^{ab}_{cd}\, ,
\ee
with
\be\la{22}
{\cal P}^{ab}_{cd} = \delta^{a}_{c}\delta^{b}_{d}-\frac{1}{N}\delta^{ab}\delta_{cd}\, .
\ee
Here, the upper indices $ab$ of the generators refer to the index for the adjoint representation and these indices are denoted by a pair of the fundamental indices. The lower indices $cd$ refer to the matrix components in the fundamental representation. For $SU(N)$, these color indices $a,b,c$ and $d$ run from $1$ to $N$.  The $N^2-N$ off-diagonal generators which correspond to $a\neq b$ are the customary ladder operators of the Cartan basis\footnote{Therefore, the order of the adjoint indices $ab$ needs to be flipped when we raise and lower these indices.}. They are normalized as
\be
\textrm{tr}(t^{ab}t^{ba})= \frac{1}{2}\, .
\ee
In the above equation, $a$ and $b$ are fixed indices and there is no summation over them. However, unlike the $N-1$ traceless diagonal generators $\lambda^d$ in the Cartan basis,
\be
\lambda^d=\frac{1}{\sqrt{2d(d+1)}}{\rm diag}(1,1,\cdots ,-d,0,0,\cdots , 0) \quad   {\rm with} \quad  d=1,2,\cdots ,N-1\, ,
\ee
there are $N$ diagonal generators in the double line basis which correspond to $a=b$ in Eq.~(\ref{21}). Therefore, this basis is overcomplete and the normalization of the diagonal generators is different from the off-diagonal ones. We have
\be
\textrm{tr}(t^{aa}t^{bb})= \frac{1}{2}\bigg(\delta^{ab}-\frac{1}{N}\bigg)\, .
\ee
As before, there is no summation over $a$ or $b$. With the help of the projection operator, the normalization for arbitrary generators is given by
\be
\textrm{tr}(t^{ab}t^{cd})= \frac{1}{2}{\cal P}^{ab,cd}=\frac{1}{2}{\cal P}^{ab}_{dc}\, ,
\ee
which also shows the orthogonality between the diagonal and off-diagonal generators.

The double line basis turns to be more useful when we consider the classical background field $A^{\rm cl}_\mu=C \delta_{\mu 0}$ as an arbitrary diagonal matrix with $(C)_{ab}=C^a \delta_{ab}$ and $\sum_{a=1}^{N} C^a=0$ for $SU(N)$. In particular, all covariant derivatives become very simple in both fundamental and adjoint representations. As a result, the computation in the presence of a background field is a trivial generalization of that in the $A_0^{{\rm cl}}=0$ case, namely, there is only a constant and color-dependent shift in the energies. For fermionic fields, the $C$-dependent momentum $P^a_\mu$ is defined as
\ba
\la{27}
P^a_\mu  =  (P^a_{0},\textbf{p})=(\tilde{\omega}_{n}+C^a,\textbf{p})\equiv ({\tilde{\omega}}^a_n,\textbf{p})\, ,
\ea
for bosonic fields, it reads
\ba
\la{28}
P^{ab}_\mu =(P^{ab}_0,\textbf{p})=(\omega_{n}+C^{ab},\textbf{p})\equiv(\omega_{n}+C^a-C^b,\textbf{p})\equiv (\omega^{ab}_n,\textbf{p})\, .
\ea
In the imaginary time formalism of the thermal field theory, the Matsubara frequencies are $\omega_{n}=2 n \pi T$ and $\tilde{\omega}_{n}=(2 n+1) \pi T-i \mu$ with $n=0,\pm1,\pm2\cdots \pm \infty$ and $\mu$ being the chemical potential. In our notations, momenta associated with a fundamental color index $a$ correspond to the fermions' momenta while the bosons' momenta are associated with an adjoint color index $ab$. With these $C$-dependent momenta, the covariant derivatives acting upon the fermionic fields $\psi$ have a simple form in momentum space, $D_\mu^{{\rm cl}} \psi_a \rightarrow -i P_\mu^a \psi_a$. Similarly, for the covariant derivative acting upon the fields in the adjoint representation, $D_\mu^{{\rm cl}} t^{ab} \rightarrow -i P_\mu^{ab} t^{ab}$.

We point out that the color structures with $A_0^{{\rm cl}}\neq 0$ is in general more complicated than those with $A_0^{{\rm cl}} = 0$. As compared to the Cartan basis, one can deal with the color structures more straightforwardly in the double line basis. It is easy to show that the product of two generators as well as the structure constants for $SU(N)$ takes a very simple form involving only the Kronecker deltas. For example,
\ba\la{29}
(t^{ab}t^{cd})_{ef}&=&\frac{1}{2}\bigg(\delta^a_e\delta^{bc}\delta^d_f-\frac{1}{N}(\delta^a_e\delta^b_f\delta^{cd}+\delta^{ab}\delta^c_e\delta^d_f)
+\frac{1}{N^2}\delta^{ab}\delta^{cd}\delta_{ef}\bigg)\, ,\nonumber\\
f^{(ab,cd,ef)}&=&\frac{i}{\sqrt{2}}\bigg(\delta^{ad}\delta^{cf}\delta^{eb}-\delta^{af}\delta^{cb}\delta^{ed}\bigg)\, .
\ea
On the other hand, in the Cartan basis, structure constants that involve both the diagonal and off-diagonal indices can not be treated in such a simple way as the above.

\begin{figure}[htbp]
\centering
\includegraphics[width=0.9\textwidth]{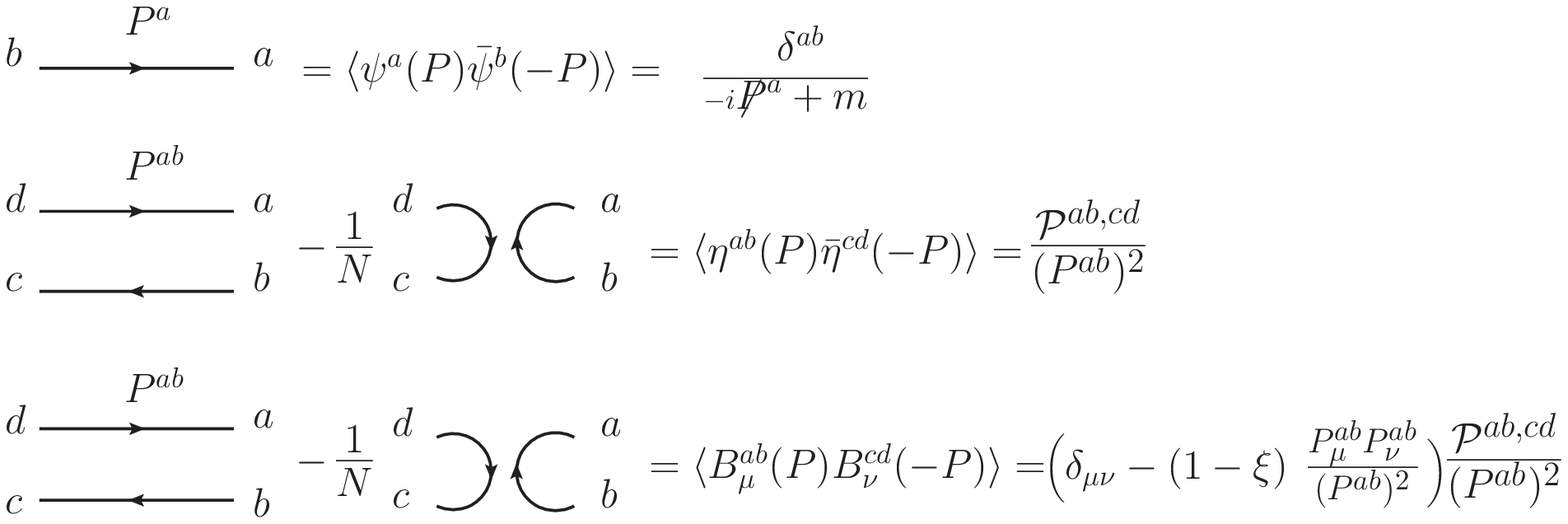}
\vspace*{-.01cm}
\caption{\small{The Feynman rules for quark, ghost and gluon propagators in double line basis.}
\label{fig1}}
\end{figure}

The corresponding Feynman rules in the double line basis can be derived straightforwardly. Adding the quark contribution $\bar{\psi} (\displaystyle{\not} D+m) \psi$ to the pure gauge action, we can obtained the explicit forms for the propagators in Fig.~\ref{fig1}.

In addition, Feynman rules for vertices are obtained by taking the derivatives of the action. The quark-gluon vertex is
\ba
-\frac{\delta {\cal S}}{\delta \psi^b(R)\delta B^{dc}_\mu(Q)\delta \bar{\psi}^a(P)}=ig(t^{cd})_{ab}\gamma_\mu\, ,
\ea
and the ghost-gluon vertex is
\ba
-\frac{\delta {\cal S}}{\delta \eta^{fe}(R)\delta B^{dc}_\mu(Q)\delta \bar{\eta}^{ba}(P)}=igf^{(ab,cd,ef)}(P^{ab})_\mu\, .
\ea
For interactions among gluons, we have the three-gluon vertex
\ba
-\frac{\delta {\cal S}}{\delta B^{fe}_\lambda(R)\delta B^{dc}_\nu(Q)\delta B^{ba}_\mu(P)}=-igf^{(ab,cd,ef)}\Gamma_{\mu\nu\lambda}(P^{ab},Q^{cd},R^{ef})\, ,
\ea
with
\ba
\Gamma_{\mu\nu\lambda}(P^{ab},Q^{cd},R^{ef})=(P^{ab}_\lambda-Q^{cd}_\lambda)\delta_{\mu\nu}+(Q^{cd}_\mu-R^{ef}_\mu)\delta_{\nu\lambda} +(R^{ef}_\nu-P^{ab}_\nu)\delta_{\lambda\mu}\, ,
\ea
and four-gluon vertex
\ba
-\frac{\delta {\cal S}}{\delta B^{hg}_\sigma(S)\delta B^{fe}_\lambda(R)\delta B^{dc}_\nu(Q)\delta B^{ba}_\mu(P)}
& = &-g^2\sum_{i,j=1}^{N}\bigg(f^{(ab,cd,ij)}f^{(ef,gh,ji)}(\delta_{\mu\lambda}\delta_{\nu\sigma}-\delta_{\mu\sigma}\delta_{\nu\lambda})\nonumber\\  &+& f^{(ab,ef,ij)}f^{(gh,cd,ji)}(\delta_{\mu\sigma}\delta_{\lambda\nu}-\delta_{\mu\nu}\delta_{\lambda\sigma})\nonumber\\ & +& f^{(ab,gh,ij)}f^{(cd,ef,ji)}(\delta_{\mu\nu}\delta_{\sigma\lambda}-\delta_{\mu\lambda}\delta_{\sigma\nu})     \bigg)\, .
\ea

In the above Feynman rules, momentum conservation applies. For example, for the three-gluon vertex, we have
\ba
P^{ab}_\mu+Q^{cd}_\mu+R^{ef}_\mu=0\, .
\ea
For the quark-gluon vertex, momentum conservation reads $P^{a}_\mu+Q^{cd}_\mu+R^{b}_\mu=0$. Notice that for anti-fermions, the direction of the momentum is opposite to that of the color flow, therefore, we have $P^{a}_0=\tilde{\omega}_n-C^a$.

\section{The two-loop QCD free energy in a constant background field}\la{fe}

\begin{figure}[htbp]
\centering
\includegraphics[width=0.9\textwidth]{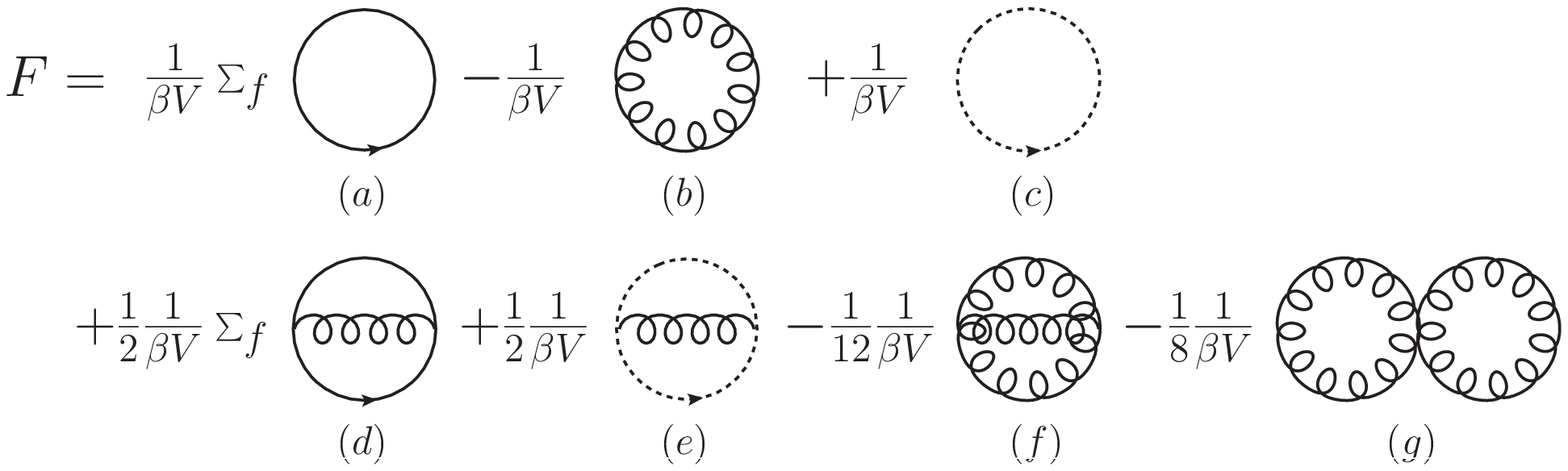}
\vspace*{-.01cm}
\caption{\small{Feynman diagrams for QCD free energy $F$ up to two-loop order.}
\label{fig2}}
\end{figure}

Consider a thermal equilibrium system in an infinite volume limit, the free energy $F$ is an important quantity to study the thermodynamic properties of the system under consideration and can be perturbatively computed from the first principle in the high temperature limit.  In this chapter, we consider $F$ in a constant background field up to two-loop order. Formally, we write $F \equiv F^{(1)}+F^{(2)}$ and the two terms correspond to the one- and two-loop contributions, respectively. Fig.\ref{fig2} shows the Feynman diagrams we need to compute. Although we will adopt the double line notation and the corresponding Feynman rules as discussed in the previous section, for simplicity, the Feynman diagrams in Fig.~\ref{fig2} are drawn in the usual manner, {\rm i.e.}, the gluon and ghost lines are not doubled.

\subsection{The one-loop free energy in a constant background field}\la{two1}

The one-loop free energy in a constant background field has been obtained in previous studies, see Refs.~\cite{Meisinger:2001cq,KorthalsAltes:1993ca,Reinosa:2014ooa,KorthalsAltes:1999cp,Reinosa:2015oua} for examples. In this sub-section, we review the calculation by introducing a new approach which is a natural generalization of that used in the $A_0^{{\rm cl}}=0$ case. With our explicit results, we will show that inclusion of a constant background field leads to a modification of the parton distribution functions while the basic structure of the free energy remains as that in vanishing background field.

We start to consider the partition function due to bosonic contribution. Utilizing the Feynman rules as introduced in Sec.~\ref{review}, it can be written as
\ba
\textrm{ln} Z^{(1)}_b=-\sum_{abcd}\sum_{n}\sum_{\textbf{q}}\textrm{ln}\{\beta^2[(\omega^{ab}_{n})^2+q^2]\}{\cal P}^{ab}_{dc}\delta_{ad}\delta_{bc}\, ,
\ea
where the subscript ``b" denotes the contribution from bosons and $\sum_{\textbf{q}} \equiv V\int\frac{d^3\textbf{q}}{(2\pi)^3}$. In the above equation, the Matsubara frequency $\omega^{ab}_{n}$ is background field-dependent which is defined in Eq.~(\ref{28}).

Carrying out the sums over the color indices $c$ and $d$, the partition function takes the following form
\ba
\la{312}
\textrm{ln} Z^{(1)}_b=-\sum_{ab}\sum_{n}\sum_{\textbf{q}}\textrm{ln}\{\beta^2[(\omega^{ab}_{n})^2+q^2]\}
\bigg(1-\frac{1}{N}\delta_{ab}\bigg)\, .
\ea

To compute the one-loop free energy $F^{(1)}_b=-\frac{\partial (T \textrm{ln} Z^{(1)}_b)}{\partial V}$, the key point is to perform the Matsubara frequency summation. At vanishing background field, this is achieved by rewriting the logarithm function in Eq.~(\ref{312}) as an integral over an auxiliary variable plus a divergent constant term which is independent on the variables of the system under consideration, such as the volume, temperature and chemical potential\cite{kapusta}. Despite a constant shift in the Matsubara frequency, we show that this approach can be generalized to the case with $A_0^{{\rm cl}}\neq 0$ where the divergent constant term is the same as before. However, the integral should be extend to the complex plane of the auxiliary variable with a proper choosing of the integration paths.

\begin{figure}[htbp]
\centering
\includegraphics[width=0.9\textwidth]{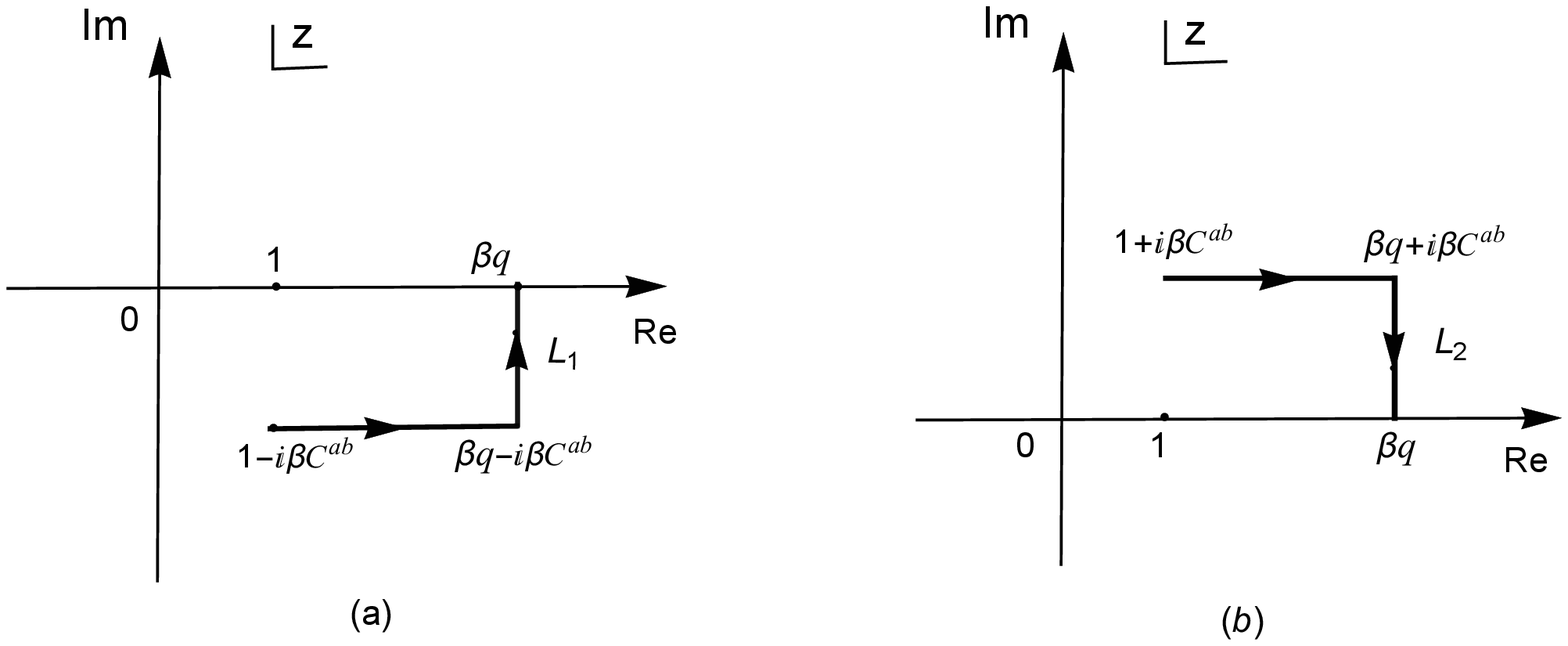}
\vspace*{-.01cm}
\caption{Integration paths for the bosonic contributions.
\label{fig:11}}
\end{figure}

First, we rewrite the partition function as a sum of the following three terms
\ba
\la{313}
\textrm{ln} Z^{(1)}_b
&=&-\sum_{ab}\sum_{n}\sum_{\textbf{q}}\bigg\{\int_{L_1}\frac{d z}{z+i(2n\pi+\beta C^{ab})}+\int_{L_2}\frac{d z}{z-i(2n\pi+\beta C^{ab})}\nonumber\\
&+&\textrm{ln}[1+(2n\pi)^2]\bigg\}\bigg(1-\frac{1}{N}\delta_{ab}\bigg)\, ,
\ea
where the paths $L_1$ and $L_2$ are shown in Fig.~\ref{fig:11} and the constant term doesn't contribute to the free energy, therefore, is dropped in the calculation.

To be more clear, consider the integrals in Eq.~(\ref{313}) with paths parallel and perpendicular to the real axis separately, we have
\ba
\la{314}
\sum_{\sigma=\pm}\int_{1}^{\beta q}\frac{d x}{(x-\sigma i\beta C^{ab})+ \sigma i(2n\pi+\beta C^{ab})}&=&\ln[(2n\pi)^2+(\beta q)^2]\nonumber\\
&-&\ln[1+(2n\pi)^2]\, .\\
\la{315}
\sum_{\sigma=\pm}\int_{\sigma \beta C^{ab}}^{0}\frac{i d y}{(\beta q+iy)-\sigma i(2n\pi+\beta C^{ab})} &=&\ln[(2n\pi+\beta C^{ab})^2+(\beta q)^2]\nonumber\\
&-&\ln[(\beta q)^2+(2n\pi)^2]\, .
\ea
Then it is easy to see the right hand side of Eq.~(\ref{313}) is equal to that of Eq.~(\ref{312}).

Interestingly, we find that integrals with paths parallel to the real axis has no dependence on the background field and is related to the free energy at $A_0^{{\rm cl}}= 0$ up to some constant term. Therefore, Eq.~(\ref{314}) eventually leads to the well-known result $-\frac{\pi^2T^4(N^2-1)}{45}$. On the other hand, the corrections due to the background field are all contained in Eq.~(\ref{315}) where the integration paths are perpendicular to the real axis.

By using the following identities\footnote{In these identities, $z$ is an arbitrary complex number, but $i z$ can not be an integer.},
\ba\la{316}
\sum_{n=-\infty}^{\infty}\frac{1}{z+i n}+\sum_{n=-\infty}^{\infty}\frac{1}{z-i n}&=&2\pi \coth(\pi z)\,,\nonumber\\
\sum_{n=-\infty}^{\infty}\frac{1}{z+i n}=\sum_{n=-\infty}^{\infty}\frac{1}{z-i n}&=&\pi \coth(\pi z)\, ,
\ea
the Matsubara frequency summation in Eq.~(\ref{313}) can be expressed in terms of the hyperbolic cotangent $\coth(z)$. After integrating over the auxiliary variable $z$, we arrive at
\ba
\textrm{ln} Z^{(1)}_b
=-\sum_{ab} V \int_0^{\infty}\frac{q^2 d q}{2\pi^2} \bigg[\beta q +\ln(1-e^{-\beta q - i\beta C^{ab}}) +\ln(1-e^{-\beta q + i\beta C^{ab}})\bigg]   \bigg(1-\frac{1}{N}\delta_{ab}\bigg)\, .\nonumber \\
\ea
In the above equation, we drop some constant terms independent on the variables of the system and these terms are exactly the same as those in the vanishing background field case. In addition, the first term in the square bracket is related to the zero-point energy which should be also removed. Then, the one-loop free energy from bosonic contribution reads
\ba
F^{(1)}_b= \sum_{ab}T\int_0^{\infty}\frac{q^2 d q}{2\pi^2}\bigg[\textrm{ln}(1-e^{-\beta q-i\beta C^{ab}})
+\textrm{ln}(1-e^{-\beta q+i\beta C^{ab}})\bigg]\bigg(1-\frac{1}{N}\delta_{ab}\bigg)\, .
\ea
Integrating by parts, we find another useful expression for the free energy which is given by
\ba
F^{(1)}_b=-\sum_{ab}\frac{1}{6\pi^2}\int_0^{\infty}q^3dq\bigg(N^{+}_{ab}(q)+N^{-}_{ab}(q)\bigg)\bigg(1-\frac{1}{N}\delta_{ab}\bigg)\, ,
\ea
where $N^{\pm}_{ab}(q)=\frac{1}{e^{\beta q \mp i\beta C^{ab}}-1}$ is the modified bosonic distribution functions in a constant background field. This equation clearly indicates that the only change due to the non-zero background field to the free energy is a redefinition of the parton distribution functions. When $A_0^{{\rm cl}}=0$, the sums over the color indices give a factor of $N^2-1$.

Integrating over the momentum $q$, the final result for the one-loop free energy can be expressed by the periodic Bernoulli polynomial $B_{4}(x)$
\ba
F^{(1)}_b&=&-\frac{T}{2\pi^2}\sum_{ab}\int_0^{\infty} q^2dq\sum_{n=1}^{\infty}\frac{1}{n}e^{-\beta q n}\bigg(e^{i\beta C^{ab}n}+e^{-i\beta C^{ab}n}\bigg)\bigg(1-\frac{1}{N}\delta_{ab}\bigg)\nonumber\\
&=&\frac{2\pi^2T^4}{3}\sum_{ab}B_{4}({\cal{C}}^{ab})+\frac{\pi^2T^4}{45}\, ,
\ea
which is reduced to $-\frac{\pi^2T^4(N^2-1)}{45}$ at vanishing background field as expected. For simplicity, we use the shorthand notation ${\cal{C}}^{ab}\equiv\frac{C^{ab}}{2\pi T}$.

It is also useful to express the above result in terms of the (trace of k-th powers of the) Polyakov loops, it reads
\ba
F^{(1)}_b=-\frac{2T^4N^2}{\pi^2}\sum_{k=1}^{\infty}\frac{1}{k^4}\ell_k \ell_k^{*}+\frac{\pi^2T^4}{45}\, ,
\ea

Next, we consider the partition function for fermions at one-loop order. We start with
\ba
\la{3111}
\textrm{ln} Z^{(1)}_f
=2N_f\sum_{d}\sum_{n}\sum_{\textbf{p}}\textrm{ln}\{\beta^2[(\tilde{\omega}_{n}^d)^2+E_P^2]\}\, ,
\ea
where $N_f$ is the flavor of fermions, the Matsubara frequency $\tilde{\omega}_{n}^d$ is defined in Eq.~(\ref{27}) and $E_P=\sqrt{p^2+m^2}$. As before, the subscript ``f" denotes the contribution from fermions.

\begin{figure}[htbp]
\centering
\includegraphics[width=0.9\textwidth]{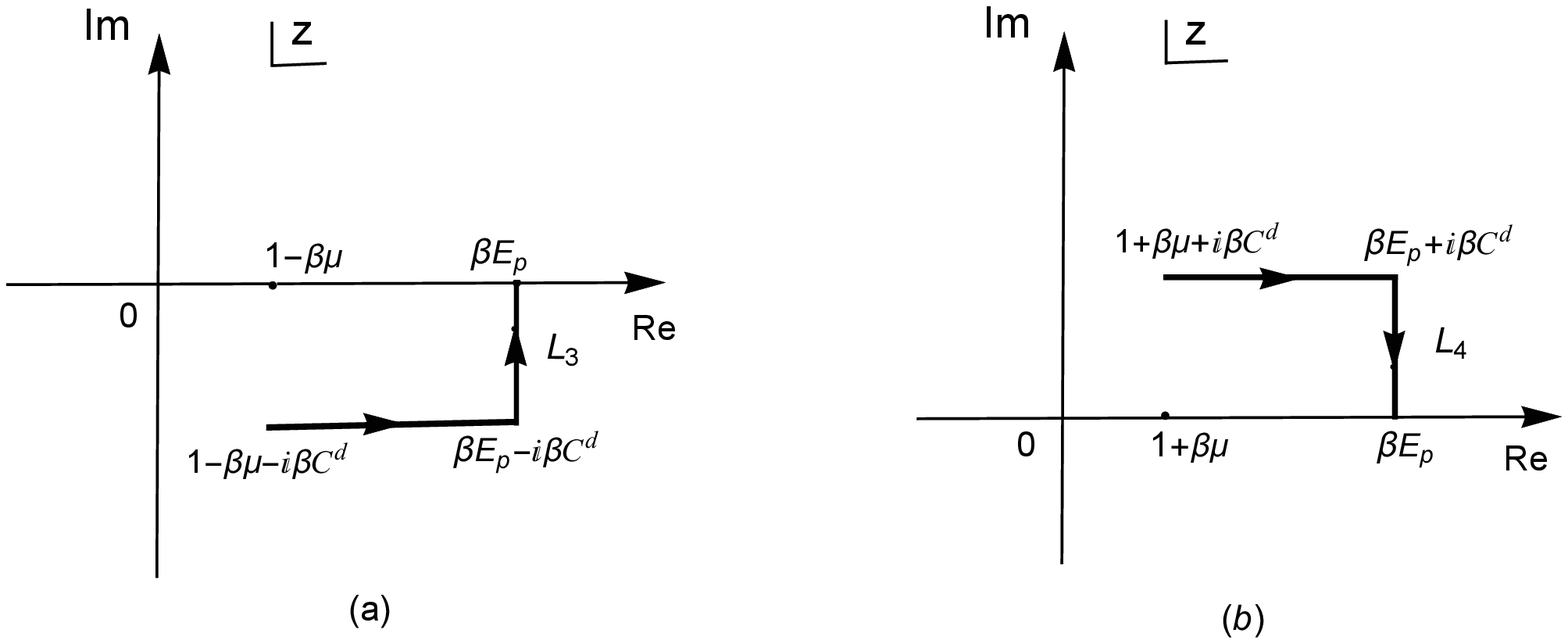}
\vspace*{-.01cm}
\caption{Integration paths for the fermionic contributions.
\label{fig:12}}
\end{figure}

The corresponding calculation for the fermions is very similar as that for bosons and Eq.~(\ref{3111}) can be rewritten as
\ba
\textrm{ln} Z^{(1)}_f
&=&2N_f\sum_{d}\sum_{n}\sum_{\textbf{p}}\bigg\{\int_{L_3}\frac{d z}{z+i(2n\pi+\pi+\beta C^{d}-i \mu \beta)}\nonumber\\&+&\int_{L_4}\frac{d z}{z-i(2n\pi+\pi+\beta C^{d}-i \mu \beta)}+\textrm{ln}[1+(2n\pi+\pi)^2]\bigg\}\, ,
\ea
where the paths $L_3$ and $L_4$ are shown in Fig.~\ref{fig:12}.

The above integrals with paths parallel to the real axis are related to the free energy at $A_0^{{\rm cl}}=0$ up to some constant term\footnote{The integrals with paths parallel to the real axis have an imaginary part which vanishes under the summation over $n$. Therefore, the free energy is real at vanishing background field.}. The corrections due to the non-zero constant background field are related to the integrals with paths perpendicular to the real axis. The rest of the calculation is straightforward and we have
\ba
\la{3113}
F^{(1)}_f=-\frac{N_f T}{\pi^2}\sum_{d}\int_0^{\infty} p^2\bigg[\textrm{ln}(1+e^{-\beta E_P-i\beta C^d-\mu \beta})+\textrm{ln}(1+e^{-\beta E_P+i\beta C^d+\mu \beta})\bigg]d p\, ,
\ea
where some constant terms as well as the zero-point energy have been removed.

The above result can be also expressed as
\ba
F^{(1)}_f=-\frac{N_f}{3\pi^2}\sum_d \int_0^{\infty}\frac{p^4}{\sqrt{p^2+m^2}}\bigg(N^{+}_{d}(p)+N^{-}_{d}(p)\bigg)d p\, ,
\ea
with $N^{\pm}_{d}(p)=\frac{1}{e^{\beta \sqrt{p^2+m^2} \mp (i\beta C^d+\beta \mu)}+1}$ being the modified fermionic distribution functions in background field\footnote{Bear in mind that in our notation the distribution function with one color index corresponds to the fermionic type, while that with two color indices corresponds to the bosonic type.}. Therefore, as the bosonic case, the one-loop fermionic free energy at $A_0^{{\rm cl}}\neq 0$ is the same as that at $A_0^{{\rm cl}}= 0$  up to a redefinition of the parton distribution functions.

In general, Eq.~(\ref{3113}) has to be evaluated numerically. On the other hand, for massless quarks, we can get an analytical expression for $F^{(1)}_f$ which is
\ba
F^{(1)}_f(m=0) = \frac{2N_f T^4}{\pi^2}\sum_d \bigg[{\rm Li}_4(- e^{- i \beta C^d-\beta \mu})+{\rm Li}_4(- e^{ i \beta C^d+\beta \mu})\bigg]\, .
\ea
In the above equation ${\rm Li}_n(z)$ is the polylogarithm function which is defined for all complex arguments $z$ with $|z| < 1$ and it can be extended to $|z|\ge 1$ by the process of analytic continuation. Using the relation between the polylogarithm functions and the Bernoulli polynomials, we can also express the above result as
\ba
F^{(1)}_f(m=0) = -\frac{4}{3} N_f T^4\pi^2 \sum_d B_4\bigg(\frac{C^d-i \mu}{2\pi T}+\frac{1}{2}\bigg)\, .
\ea
The one-loop fermionic free energy is complex with non-zero $C^d$ and $\mu$. Using Eq.~(\ref{bercpl}), we can write down the explicit results for the real and imaginary part of the free energy $F^{(1)}_f$ in the massless limit as the following
\ba
\la{3117}
{\rm Re} \, F_f^{(1)}\big|_{m=0} &=& -\frac{N_f T^4}{12\pi^2}\sum_d \bigg[\bigg(\frac{\mu}{T}\bigg)^4-24\pi^2 B_2({\cal{C}}^d)\bigg(\frac{\mu}{T}\bigg)^2+16 \pi^4B_4({\cal{C}}^d)\bigg]\, ,\nonumber\\
{\rm Im} \, F_f^{(1)}\big|_{m=0} &=& -\frac{2N_f T^4}{3\pi}\sum_d \bigg[B_1({\cal{C}}^d)\bigg(\frac{\mu}{T}\bigg)^3-4 \pi^2B_3({\cal{C}}^d)\bigg(\frac{\mu}{T}\bigg)\bigg]\, ,
\ea
where the shorthand notation ${\cal{C}}^d \equiv \frac{C^d}{2\pi T}+\frac{1}{2}$. Notice that there is an extra factor $\frac{1}{2}$ as compared to the corresponding definition  ${\cal{C}}^{ab}\equiv\frac{C^{ab}}{2\pi T}$ for bosons. Similarly, we can also rewrite the above results in terms of the Polyakov loops.
\ba
\la{3117add}
{\rm Re} \, F_f^{(1)}\big|_{m=0} &=& -\frac{N_f N T^4}{12\pi^2}\bigg\{\bigg(\frac{\mu}{T}\bigg)^4-24\sum_{k=1}^{\infty}\bigg[\bigg(\frac{\mu}{T}\bigg)^2+\bigg(\frac{2}{k^2}\bigg)\bigg]\frac{(-1)^k}{k^2}
{\rm Re}\,\ell_k\bigg\}\, ,\nonumber\\
{\rm Im} \, F_f^{(1)}\big|_{m=0} &=& \frac{2N_f N T^4}{3\pi^2} \bigg(\frac{\mu}{T}\bigg)\sum_{k=1}^{\infty}\bigg[\bigg(\frac{\mu}{T}\bigg)^2+\bigg(\frac{6}{k^2}\bigg)\bigg]\frac{(-1)^k}{k}{\rm Im}\,\ell_k\, ,
\ea

The background fields are arbitrary in the above equations. However, for a special case where $\frac{C^d}{2\pi T}=\frac{N-2d+1}{2N}s$ (with $0\le s\le 1$), the imaginary part of the free energy vanishes. This parametrization of the background fields is known as the straight line {\em ansatz} which has been used in Ref.~\cite{Dumitru:2012fw} to study the deconfining phase transition. The absent of the imaginary part in this {\em ansatz} is actually true for general situation with finite quark mass. This is clear when we formally rewritten Eq.~(\ref{3113}) as an infinite sum of the modified Bessel function of the second kind $K_2(x)$,
\ba
F^{(1)}_f=- N_f\frac{T^2m^2}{\pi^2}\sum_d\sum_{n=1}^{\infty}\frac{(-1)^{n-1}}{n^2}\bigg(e^{i\beta C^d n+\beta\mu n}+e^{-i\beta C^d n-\beta\mu n}\bigg)K_2[\beta m n]\, .
\ea

On the other hand, at zero background field, the free energy $F^{(1)}_f$ becomes real and it is easy to check that the above equation is reduced to the well-known result
\ba
F^{(1)}_f(m=0,C=0) = -\frac{7\pi^2 N N_f T^4}{180}\bigg[1+\frac{30}{7\pi^2}\bigg(\frac{\mu}{T}\bigg)^2+\frac{15}{7\pi^4}\bigg(\frac{\mu}{T}\bigg)^4\bigg]\, .
\ea

\subsection{The two-loop free energy of bosons} \la{two2}

Now we consider the last three diagrams in Fig.~\ref{fig2}  that contribute to the two-loop free energy of bosons. We employ the double line notation and explicitly show that the bosonic free energy at two-loop with $A_0^{{\rm cl}}\neq 0$ is similar as that with $A_0^{{\rm cl}} = 0$ up to a redefinition of the parton distributions functions. This is exactly the same conclusion as the one-loop case. It turns out the corresponding calculations can be reduced to the bosonic sum-integral which is computed in Appendix \ref{appendixB}.

Using the Feynman rules in double line basis, we can write down the expressions for the three diagrams in Fig.~\ref{fig2}. Take diagram $(e)$ as an example, we have\footnote{In this subsection, we use the Feynman gauge by setting $\xi=1$. The gauge-dependence of the free energy will be discussed latter.}
\ba
F^{(e)} &=&-\frac{g^2}{2}\sum_{{\rm colors}}\sumint\frac{d^4P}{(2\pi)^4}\frac{d^4K}{(2\pi)^4}\frac{d^4Q}{(2\pi)^4}(2\pi)^4\delta^{4}(Q^{\bar{a}\bar{b}}+K^{ab}-P^{a' b'})\nonumber\\ &\times & \frac{{\cal P}^{ab,cd}{\cal P}^{a'b',c'd'}{\cal P}^{\bar{a}\bar{b},\bar{c}\bar{d}}}{(K^{ab})^2(P^{a'b'})^2(Q^{\bar{a}\bar{b}})^2}(P^{d'c'}\cdot K^{dc})f^{(c'd',\bar{a}\bar{b},ab)}f^{(cd,\bar{c}\bar{d},a'b')}\, .
\ea
Here, the superscript $(e)$ denotes to the corresponding diagram in Fig.~\ref{fig1} and the sum-integral $\sumint\frac{d^4P}{(2\pi)^4}\equiv  T \sum_{n_p} \int\frac{d^3 {\bf{p}}}{(2\pi)^3}$ and $n_p$ is the Matsubara frequency for momentum $P$. As we can see, the color structure becomes complicated when considering the non-zero background field. However, the corresponding sums can be done straightforwardly by using Eqs.~(\ref{22}) and (\ref{29}) and we arrive at
\ba
\la{322}
& &F^{(e)} = \frac{g^2}{4}\sum_{abc}\sumint\frac{d^{4}P}{(2\pi)^{4}}\frac{d^{4}K}{(2\pi)^{4}}\frac{d^{4}Q}{(2\pi)^{4}}\bigg\{ \delta^4(Q^{cb}-P^{ca}+K^{ba})\frac{P^{ca}\cdot K^{ba}}{(K^{ba})^2(P^{ca})^2(Q^{cb})^2}\nonumber\\ & &+\delta^4(Q^{cb}-P^{ab}+K^{ac})\frac{P^{ab}\cdot K^{ac}}{(K^{ac})^2(P^{ab})^2(Q^{cb})^2}\bigg\}\bigg(1-\delta_{bc}\delta_{ac}\delta_{ba}\bigg)\, .
\ea
It is very important to find that in the above equation the sum-integrals can be performed simultaneously and independently, therefore, Eq.~(\ref{322}) can be simplified into a product of two independent bosonic sum-integrals as the following
\ba
\la{323}
F^{(e)}  &=&  \frac{g^2}{8}\sum_{abc}\sumint\frac{d^{4}K}{(2\pi)^{4}}\frac{d^{4}Q}{(2\pi)^{4}}\bigg\{ \frac{1}{(K^{ba})^2(Q^{cb})^2}+\frac{1}{(K^{ac})^2(Q^{cb})^2}\bigg\}\bigg(1-\delta_{bc}\delta_{ac}\delta_{ba}\bigg)\nonumber\\
&=& \frac{g^2}{4}\sum_{abc}\sumint\frac{d^{4}K}{(2\pi)^{4}}\frac{d^{4}Q}{(2\pi)^{4}} \frac{1}{(K^{ab})^2(Q^{bc})^2}\bigg(1-\delta_{bc}\delta_{ac}\delta_{ab}\bigg)\,.
\ea
In the second line of the above equation, we used the fact that bosonic sum-integral is invariant under the exchange of the color indices.

Using Eq.~(\ref{ab2}), we can rewrite Eq.~(\ref{323}) in terms of the parton distribution functions
\ba
F^{(e)}  =\frac{g^2}{64\pi^4}\sum_{abc}\bigg(\int q  N_{ab}(q) dq \bigg) \bigg( \int k  N_{bc}(k)dk\bigg )\bigg(1-\delta_{bc}\delta_{ac}\delta_{ab}\bigg)  ,
\ea
where $N_{ab}(q) \equiv N_{ab}^{+}(q)+N_{ab}^{-}(q)$ and the constant term in Eq.~(\ref{ab2}) has been dropped. Notice that the above expression has the same structure as the corresponding contribution at vanishing background field. In the latter case, $\sum_{abc}(1-\delta_{bc}\delta_{ac}\delta_{ba})=N(N^2-1)$.

After carrying out the integrals over the momenta, according to Eq.~(\ref{ab5}), the final result can be expressed by the periodic Bernoulli polynomial $B_2(x)$ as
\ba
F^{(e)}  =\frac{g^2T^4}{16}\sum_{abc}B_2({\cal{C}}^{ba})B_2({\cal{C}}^{cb})-\frac{g^2NT^4}{576}\, .
\ea

The calculations of the remaining two diagrams in Fig.~\ref{fig2} are tedious but straightforward by using the same method as discussed above. The results turn to be very simple, each of the three diagrams shares the same structure as the second line of Eq.~(\ref{323}) and the only change is the prefactor. Instead of $1/4$, we get $-9/4$ for diagram $(f)$ and $3$ for diagram $(g)$. Summing up the contributions from all the three diagrams, the bosonic free energy at two-loop order reads
\ba
\la{326}
F^{(2)}_b &=&  g^2\sum_{abc}\sumint\frac{d^{4}K}{(2\pi)^{4}}\frac{d^{4}Q}{(2\pi)^{4}} \frac{1}{(K^{ab})^2(Q^{bc})^2}\bigg(1-\delta_{bc}\delta_{ac}\delta_{ab}\bigg)\nonumber\\ &= &\frac{g^2}{16\pi^4}\sum_{abc}\bigg(\int q  N_{ab}(q) dq \bigg) \bigg( \int k  N_{bc}(k)dk\bigg )\bigg(1-\delta_{bc}\delta_{ac}\delta_{ab}\bigg)\nonumber\\ &=&\frac{g^2T^4}{4}\sum_{abc}B_2({\cal{C}}^{ab})B_2({\cal{C}}^{bc})-\frac{g^2NT^4}{144}\, .
\ea
It is easy to verify that when the background fields vanish, the potential equals to $\frac{g^2T^4(N^3-N)}{144}$ as expected.

\subsection{The two-loop free energy of fermions}\la{two3}

The diagram $(d)$ in Fig.~\ref{fig2} contributes to the two-loop free energy of fermions. According to the Feynman rules, we get
\ba
\la{331}
F_f^{(2)} &=& \frac{g^2}{2}N_f\sum_{{\rm colors}}\sumint \frac{d^4P}{(2\pi)^4}\frac{d^4K}{(2\pi)^4}\frac{d^4Q}{(2\pi)^4}(2\pi)^4\delta^4(P^c-K^a+Q^{a'b'})\nonumber\\&\times&\textrm{tr}
\bigg[\gamma^{\mu}(\displaystyle{\not}K^a-i m)
\gamma^{\nu}(\displaystyle{\not}P^c-i m)\bigg]\frac{(t^{c'd'})_{da}\delta^{ab}}{(K^a)^2+m^2}\frac{(t^{a'b'})_{bc}\delta^{cd}}{(P^c)^2+m^2}\frac{{\cal P}^{a'b',c'd'}}{(Q^{a'b'})^2}\delta_{\mu\nu}\, .
\ea
After sum over all the color indices except $b$ and $d$ and carry out the trace, the above equation becomes
\ba
\la{332}
F_f^{(2)} &=&-2 g^2N_f\sum_{db}\sumint \frac{d^4P}{(2\pi)^4}\frac{d^4K}{(2\pi)^4}\frac{d^4Q}{(2\pi)^4}(2\pi)^4\delta^4(P^d-K^b+Q^{bd})\nonumber\\
&\times&\frac{P^d\cdot K^b+2m^2}{[(K^{b})^2+m^2][(P^d)^2+m^2](Q^{bd})^2}\bigg(1-\frac{1}{N}\delta_{bd}\bigg)\, .
\ea
Unlike the bosonic case, the above sum-integrals are not independent, therefore, can not be simply carried out by using the formulas as given in Appendix~\ref{appendixB}. This actually makes the corresponding calculation much more difficulty as compared to the bosonic free energy. At $A_0^{{\rm cl}}=0$, traditional approach is to rewrite the Kronecker delta $\delta_{n_k,n_q+n_p}$ in terms of the exponential functions,
 then each Matsubara frequency sum is converted to a contour integral and the three contour integrations can be performed independently and simultaneously. We find that such a method can be generalized to $A_0^{{\rm cl}}\neq 0$ and the bosonic and fermionic sums are given by
\be
\la{333}
T\sum_{n_q}\frac{I(P^d_0,K^b_0,Q^{bd}_0)}{(Q^{bd}_0)^2+q^2}=\frac{I(P^d_0,K^b_0,i q)}{2 q}N^-_{bd}(q)+\frac{I(P^d_0,K^b_0,-i
q)}{2 q}(N^+_{bd}(q)+1)\, ,
\ee
and
\be
\la{334}
T\sum_{n_p}\frac{I(P^d_0,K^b_0,Q^{bd}_0)}{(P^d_0)^2+E_p^2}=\frac{I(i E_p,K^b_0,Q^{bd}_0)}{2E_p}(1-N^-_{d}(p))-\frac{I(-i E_p,K^b_0,Q^{bd}_0)}{2E_p}N^+_{d}(p)\, ,
\ee
where $I(P^d_0,K^b_0,Q^{bd}_0)$ is a function which has no singularities on the complex plane and its explicit form reads
\be\la{335}
I(P^d_0,K^b_0,Q^{bd}_0)=\frac{P^d_0\cdot K^b_0+{\bf {p}}\cdot{\bf {k}}+2m^2}{i(K^b_0-P^d_0-Q^{bd}_0)}[e^{i\beta(P^d_0+i\mu-C^d)}-e^{i\beta(K^b_0-Q^{bd}_0+i\mu-C^d)}]\, .
\ee

With the above formulas, remaining calculations can be carried out without any technical difficulty. However, to arrive at the final result, one needs to do a rather tedious algebra which we don't show the details here. Instead, we consider an alternative way to do the Matsubara frequency sums which turns to be relatively simpler. 

The first step is to rewrite Eq.~(\ref{332}) as the following
\ba
\la{336}
F_f^{(2)} &=& -g^2N_f\sum_{db}\sumint \frac{d^4P}{(2\pi)^4}\frac{d^4K}{(2\pi)^4}\frac{d^4Q}{(2\pi)^4}\frac{2m^2(2\pi)^4\delta^4(P^d-K^b+Q^{bd})(1-\frac{1}{N}\delta_{bd})}
{[(K^{b})^2+m^2][(P^{d})^2+m^2](Q^{bd})^2}\nonumber\\
&+& g^2N_f\sum_{db}\sumint \frac{d^4K}{(2\pi)^4}\frac{d^4P}{(2\pi)^4}\frac{1-\frac{1}{N}\delta_{bd}}{[(K^{b})^2+m^2][(P^{d})^2+m^2]}\nonumber\\
&-&g^2N_f\sum_{db}\sumint \frac{d^4K}{(2\pi)^4}\frac{d^4Q}{(2\pi)^4}\frac{1-\frac{1}{N}\delta_{bd}}{[(K^{b})^2+m^2](Q^{bd})^2}\nonumber\\
&-&g^2N_f\sum_{db}\sumint \frac{d^4K}{(2\pi)^4}\frac{d^4Q}{(2\pi)^4}\frac{1-\frac{1}{N}\delta_{bd}}{[(K^{d})^2+m^2](Q^{bd})^2}\,.
\ea
At first glance, the above equation looks more complicated than the original one. However, it is actually a proper arrangement of Eq.~(\ref{332}). For the last three lines in Eq.~(\ref{336}), the delta function is eliminated directly by doing one sum-integral and the remaining two sum-integrals can be simply computed by using the sum-integral formulas as given in Appendix \ref{appendixB} because $\sumint d^4K$ and $\sumint d^4P$(or $\sumint d^4Q$) are independent there. In terms of the parton distribution functions, the result from the three lines in Eq.~(\ref{336}) is given by
\ba
\la{337}
\frac{g^2N_f}{16\pi^4}\sum_{db}&&\bigg[\bigg(\int \frac{p^2}{E_p}  N_{d}(p) dp \bigg)\bigg(\int \frac{k^2}{E_k}  N_{b}(k) dk \bigg)+\bigg(\int \frac{k^2}{E_k}  (N_{b}(k)+N_{d}(k)) dk \bigg)\nonumber \\ &\times& \bigg(\int q  N_{bd}(q) dq \bigg)\bigg]\bigg(1-\frac{1}{N}\delta_{bd}\bigg)\,,
\ea
where terms that are linear in or independent on the distribution function are dropped. Those that are independent correspond to the vacuum energy shift while those that are linear are canceled by the parton vacuum self-energy renormalizations. In addition, we define the fermionic distribution function $N_b(p)\equiv N^+_b(p)+ N^-_b(p)$.

However, this is not the case for the first line in Eq.~(\ref{336}) where the sum-integrals can not be carried out simultaneously as above. Obviously, this contribution vanishes in the zero quark mass limit. Therefore, we will first consider the two-loop free energy for massless fermions. In this limit, integrations in Eq.~(\ref{337}) can be done analytically. In general, the free energy is complex and its real part reads
\ba\la{338}
{\rm Re} \, F_f^{(2)}\big|_{m=0} &=& \frac{g^2N_fT^4}{4}\sum_{db}\bigg\{\frac{1}{16\pi^4}\bigg(\frac{\mu}{T}\bigg)^4-\frac{1}{2\pi^2}\bigg(\frac{\mu}{T}\bigg)^2\bigg[B_2({\cal{C}}^b)- B_2({\cal{C}}^{bd})
\nonumber \\&+&2 B_1({\cal{C}}^b)B_1({\cal{C}}^d)\bigg]
+B_2({\cal{C}}^b)\bigg[B_2({\cal{C}}^d)-2B_2({\cal{C}}^{bd})\bigg]
\bigg\}\bigg(1-\frac{1}{N}\delta_{bd}\bigg)\,,\nonumber \\
\ea
while the imaginary part is given by
\ba\la{339}
{\rm Im} \, F_f^{(2)}\big|_{m=0} &=& -\frac{g^2N_fT^4}{2\pi}\frac{\mu}{T}\sum_{db}\bigg\{B_1({\cal{C}}^b)\bigg[B_2({\cal{C}}^d)-B_2( {\cal{C}}^{bd})\bigg]\nonumber \\
&-&\frac{1}{4\pi^2}\bigg(\frac{\mu}{T}\bigg)^2 B_1({\cal{C}}^b)\bigg\}\bigg(1-\frac{1}{N}\delta_{bd}\bigg)\,.
\ea
Using the properties of Bernoulli polynomials, one can check that the imaginary part vanishes under the summation over the color indices if the straight line {\em ansatz} of the background field is satisfied. In addition, at vanishing background field, the free energy becomes real as expected and we obtain the well-known result
\ba
F_f^{(2)}(m=0,C=0)|_{\rm Re} = \frac{5 g^2N_fT^4}{576}(N^2-1)\bigg[1+\frac{18}{5\pi^2}\bigg(\frac{\mu}{T}\bigg)^2+\frac{9}{5\pi^4}\bigg(\frac{\mu}{T}\bigg)^4\bigg]\,.
\ea

So far, the calculation of the two-loop free energy doesn't show any complication as compared to the one-loop calculation. To be more specifically, all the Matsubara frequency sums can be easily carried out by using the basic identity as given in Eq.~(\ref{316}). Now we turn to calculation of the first line in Eq.~(\ref{336}). Although the sum-integrals are dependent on each other, with a careful inspection, the basic formula that we need in the frequency sums is actually simple and can be obtained directly from Eq.~(\ref{316}). It reads
\be\la{3311}
\sum_n\frac{1}{(n+z_1)(n+z_2)}=\frac{\pi}{z_2-z_1}[\cot(\pi z_1)-\cot(\pi z_2)]\, .
\ee
With the help of the above equation, the fermionic free energy can be computed more efficiently than doing contour integrals as done in the traditional way. In fact, we find that under the integrations over the three-momenta, different terms in the frequency sum contribute equally to the free energy. Therefore, it is not necessary to get the exact result of the Matsubara frequency sums and the corresponding calculation can be significantly simplified. For more details, one can refer to Appendix~\ref{appendixC}.

Using Eq.~(\ref{ac15}), the Matsubara frequency sums in the first line in Eq.~(\ref{336}) can be obtained. Together with Eq.~(\ref{337}), we can get the full result of the two-loop fermionic free energy with finite quark mass and chemical potential. For latter convenience, we formally divided it into the following four parts,
\ba
\la{3312}
F_f^{(2)}=F_f^{(2a)}+F_f^{(2b)}+F_f^{(2c)}+F_f^{(2d)}\, ,
\ea
with
\ba
F_f^{(2a)}&=&\frac{g^2N_f}{4}\sum_{db}\bigg(1-\frac{1}{N}\delta_{bd}\bigg)\bigg\{\int_{\textbf{k}}2T^2 B_2( {\cal {C}}^{bd})\frac{N_b(k)}{E_k}+\int_{\textbf{k}}\int_{\textbf{p}}\frac{1}{E_k E_p}\nonumber\\
&\times&\bigg[\bigg(N^+_{b}(k)N^+_{d}(p)+N^-_{b}(k)N^-_{d}(p)\bigg)\bigg(1+ \frac{2m^2}{(E_p-E_k)^2-(\textbf{p}-\textbf{k})^2}\bigg)\nonumber\\& +&\bigg(N^+_{b}(k)N^-_{d}(p)+N^-_{b}(k)N^+_{d}(p)\bigg)\bigg(1+\frac{2m^2}{(E_p+E_k)^2-(\textbf{p}-\textbf{k})^2}\bigg)\bigg]\bigg\}\, ,
\ea
\ba\la{3314}
F_f^{(2b)}  &=&\frac{g^2N_f}{4}\sum_{db}\int_{\textbf{p}}\int_{\textbf{k}}\frac{2m^2}{(m^2+\textbf{p}\cdot\textbf{k})^2-E_p^2E_k^2}
\bar{N}_{bd}(|\textbf{p}-\textbf{k}|)
\bar{N}_{d}(p)\nonumber\\
&=&i\frac{g^2m^2N_fT}{4\pi}\sum_{db}\int_{\textbf{p}}\frac{\bar{N}_{d}(p)}{p E_p}{\rm ln}\bigg[\frac{E_p+p}{E_p-p}\bigg]B_1({\cal {C}}^{bd})
\, ,
\ea
\ba
F_f^{(2c)} &=&\frac{g^2N_f}{2}\sum_{db}\bigg(1-\frac{1}{N}\delta_{bd}\bigg)\int_{\textbf{p}}\int_{\textbf{k}}\bigg\{\frac{N_{b}(k)}
{E_p E_ k|\textbf{p}-\textbf{k}|} \bigg[\frac{2m^2+E_p E_k+\textbf{p}\cdot\textbf{k}}{E_p+E_k+|\textbf{p}-\textbf{k}|} \nonumber\\&+&\frac{2m^2-E_p E_k+\textbf{p}\cdot\textbf{k}}{E_p-E_k+|\textbf{p}-\textbf{k}|}\bigg]-\frac{N_{bd}(|\textbf{p}-\textbf{k}|)}{E_k |\textbf{p}-\textbf{k}|}\bigg[1+\frac{2m^2}{(E_p+E_k)^2-(\textbf{p}-\textbf{k})^2}\bigg]\bigg\} \, ,
\ea
\ba\la{3316}
F_f^{(2d)} &=&-\frac{g^2 N_f}{2}(N^2-1)\int_{\textbf{p}}\int_{\textbf{k}}\frac{1}{E_p E_k|\textbf{p}-\textbf{k}|}\frac{2m^2+E_p E_k+\textbf{p}\cdot\textbf{k}}{E_p+E_k+|\textbf{p}-\textbf{k}|}\, .
\ea
In the above equations, we use $\int_{\textbf{q}}\equiv \int\frac{d^3 \textbf{q}}{(2\pi)^3}$. Several remarks are in order. The $F_f^{(2a)}$ and $F_f^{(2b)}$ terms are quadratic in the parton distribution functions. $F_f^{(2a)}$ is familiar to us since it has the same functional structure as the fermionic free energy at vanishing background field. The only change is a redefinition of the distribution function as the bosonic case. $F_f^{(2b)}$ is new. This term is related to the difference of the distribution functions which is denoted as $\bar{N}_{bd}(p)\equiv N^+_{bd}(p)-N^-_{bd}(p)$ and $\bar{N}_{b}(k)\equiv N^+_{b}(k)-N^-_{b}(k)$. It disappears when $A_0^{{\rm cl}}=0$ by definition and is also absent in the massless case. $F_f^{(2c)}$ is linear in the distribution function and is canceled by the fermion/boson vacuum self-energy renormalizations\cite{kapusta}. The renormalization procedure is a trivial generalization of that with $A_0^{{\rm cl}}=0$ because $F_f^{(2c)}$ also has the same functional structure as the corresponding form at vanishing background field. $F_f^{(2d)}$ is exactly the same as that with $A_0^{{\rm cl}}=0$ because it is independent on the temperature which is just an energy shift of the vacuum and not of interest here.

\subsection{Gauge dependence of the free energy} \la{two3}

As we know that at vanishing background field, the two-loop free energy is gauge independent. 
Although each of the gluonic contribution is dependent on the gauge parameter $\xi$, the sum of the three diagrams is not dependent on $\xi$. However, this is no longer true when the background field is considered. Both the fermionic and gluonic contributions become gauge dependent. In this section, we give the explicit expressions of the gauge dependent part for each diagram in Fig.~\ref{fig1}. In the last section, we choose the gauge parameter $\xi=1$, therefore, for arbitrary $\xi$, the remaining contributions could be linear, quadratic or cubic in $1-\xi$.

Using the Feynman rules, the remaining contributions from each diagram are given by
\ba\la{341}
F^{(e)}(\xi) &=&-\frac{g^2}{8}(1-\xi)\sum_{abc} \sumint \frac{d^4P}{(2\pi)^4}\frac{d^4Q}{(2\pi)^4}\bigg\{\frac{1}{(Q^{ca})^2(P^{cb})^2}+\frac{4( Q^{ca}\cdot P^{ab})}{(Q^{ca})^4(P^{ab})^2}\bigg\}\bigg(1-\delta_{ab}\delta_{bc}\delta_{ac}\bigg)\nonumber \\
F^{(f)}(\xi) &=&-\frac{g^2}{8}(1-\xi)\sum_{abc} \sumint \frac{d^4P}{(2\pi)^4}\frac{d^4Q}{(2\pi)^4}\bigg\{2(1-\xi)\bigg(\frac{1}{(Q^{ca})^2(P^{cb})^2}-\frac{(Q^{ca}\cdot P^{ab})^2}{(Q^{ca})^4(P^{ab})^2}\bigg)\nonumber \\
&-&\frac{13}{(Q^{ca})^2(P^{cb})^2}-\frac{20(Q^{ca}\cdot P^{ab})}{(Q^{ca})^4(P^{ab})^2}\bigg\}\bigg(1-\delta_{ab}\delta_{bc}\delta_{ac}\bigg)\nonumber \\
F^{(g)}(\xi) &=&-\frac{g^2}{8}(1-\xi)\sum_{abc} \sumint \frac{d^4P}{(2\pi)^4}\frac{d^4Q}{(2\pi)^4}\bigg\{2(1-\xi)\bigg(\frac{(Q^{ca}\cdot P^{ab})^2}{(Q^{ca})^4(P^{ab})^2}-\frac{1}{(Q^{ca})^2(P^{cb})^2}\bigg)\nonumber \\
&+&\frac{12}{(Q^{ca})^2(P^{cb})^2}\bigg\}\bigg(1-\delta_{ab}\delta_{bc}\delta_{ac}\bigg)\,.
\ea
In the above equations, the structure $\frac{ Q^{ca}\cdot P^{ab}}{(Q^{ca})^4(P^{ab})^2}$ is new for non-vanishing background field and has finite contribution to the free energy. On the other hand, the other contributions are zero because of the cancelation among the three diagrams. Such a cancelation has exactly the same manner as that happens at vanishing background field.

As a result, the sum of the three diagrams is simple and the two sum-integrals we need to compute can be considered independently. Using the basic formulas as given in Eqs.~(\ref{ab6}) and (\ref{ab8}), the final result can also be expressed in terms of the periodic Bernoulli polynomials. Notice that $Q^{ca}\cdot P^{ab}=Q_0^{ca}\cdot P_0^{ab}+{\bf q}\cdot {\bf p}$ and the term $\frac{{\bf q}\cdot {\bf p}}{(Q^{ca})^4(P^{ab})^2}$ vanishes after we integrate over the three-momentum. Therefore, we have
\ba\la{342}
F^{(2)}_b(\xi) &=&2g^2(1-\xi)\sum_{abc} \sumint \frac{d^4P}{(2\pi)^4}\frac{d^4Q}{(2\pi)^4}\bigg(\frac{Q_0^{ca}\cdot P_0^{ab}}{(Q^{ca})^4(P^{ab})^2}\bigg)\nonumber \\&=&2g^2(1-\xi)\sum_{abc} \bigg(\sumint \frac{d^4P}{(2\pi)^4}\frac{P_0^{ab}}{(P^{ab})^2}\bigg)\bigg(\sumint\frac{d^4Q}{(2\pi)^4}\frac{Q_0^{ca}}{(Q^{ca})^4}\bigg)\nonumber \\
&=&-\frac{g^2 T^4}{3}(1-\xi)\sum_{abc} B_3({\cal{C}}^{ba}) B_1({\cal{C}}^{ac})\,.
\ea

Similarly, we can get the remaining gauge dependent part for the fermionic contribution as shown by diagram (d) in Fig.~\ref{fig1},
\ba
\la{343}
F^{(2)}_f(\xi) &=&2g^2N_f(1-\xi)\sum_{ab} \bigg(\sumint \frac{d^4P}{(2\pi)^4}\frac{P_0^{a}}{(P^{a})^2+m^2}\bigg)\bigg(\sumint\frac{d^4Q}{(2\pi)^4}\frac{Q_0^{ab}}{(Q^{ab})^4}\bigg)\nonumber \\
&=&-\frac{g^2N_f T}{2\pi}(1-\xi)\sum_{ab} \bigg(\sumint \frac{d^4P}{(2\pi)^4}\frac{P_0^{a}}{(P^{a})^2+m^2}\bigg)B_1({\cal{C}}^{ab})\nonumber \\
&\xrightarrow{m=0}&-\frac{g^2N_f T^4}{3}(1-\xi)\sum_{ab} B_3\bigg(\frac{C^{a}-i \mu}{2\pi T}+\frac{1}{2} \bigg)B_1({\cal{C}}^{ab})\, ,
\ea
where Eq.~(\ref{ab7}) is used for the fermionic sum-integral. Although it is not obvious, our calculation shows that the two sum-integrals in $F^{(2)}_f(\xi)$ are not dependent on each other which makes the calculation much simpler than Eq.~(\ref{332}). We mention that Eqs.~(\ref{342}) and (\ref{343}) contain the distribution functions $\bar{N}_a(p)$ and $\bar{N}_{ab}(q)$ which vanish by definition when $A_0^{{\rm cl}}=0$. Therefore, both equations only exist when a non-zero background field is taken into account.
\section{The insertion contribution and cancellation of the gauge dependence}\la{in}

For the constrained version of the effective potential, we have introduced an extra field $\epsilon$. So far, our calculation has nothing to do with this field. In fact, at one-loop order, when one expands the constrained action to quadratic order of the quantum fluctuations which is formally denoted by $S_{{\rm con}}^{(2)}$, the extra field $\epsilon$ already appears. However, the integral over the quantum fluctuation $\epsilon_q$ gives only the delta constraints on the $N-1$ variables $\hat{Q}_0^n\equiv{\rm Tr}[Q_0(0) ({\bf L}^n(C))^\prime]$ with $n=1,2,\cdots, N-1$. Here, $Q_0(0)$ is the quantum fluctuation with zero-momentum which is defined as $\int d^3\vec x ~ d\tau\, Q_0(\tau, \vec{x})/V$. As argued in Ref.~\cite{KorthalsAltes:1993ca}, these $N-1$ constraints won't change the thermodynamics in the large volume limit, therefore, the one-loop effective potential coincides with the free energy as we have computed before, namely $\Gamma^{(1)}=F^{(1)}$ for both bosons and fermions.

For the two-loop contributions, we need to expand the constrained action to order $g^2$. Apart from the terms that give rise to the QCD free energy, a non-trivial term related to the extra field $\epsilon$ appears in $(S_{{\rm con}}^{(3)})^2$. As compared to the one-loop case, there is an extra factor linear in the quantum fluctuation $\epsilon_q$ appearing in the functional integral,
\be
(S_{{\rm con}}^{(3)})^2=g^2 i\epsilon_q \; \Tr [\overline{Q_0^2}\cdot {\bf L}''(C)] \;
\left( Q^3\cdot S'''(C) \right)\, ,
\ee
where the bar means the average over the volume $V$ and terms that don't contribute at ${\cal O}(V)$ are neglected. We also introduce the shorthand notation ``$\cdot$"  to indicate integrations over space time and summations over the Lorentz and color indices. Notice that the above expression is exact for $SU(2)$ and the generalization to $SU(N)$ is straightforward. One can check that doing the integration over $\epsilon_q$ will lead to a derivative $\partial/\partial{\hat{Q}_0}$ acting on the terms in the expansion of the action. Therefore, at two-loop order, such a non-trivial term leads to a new contribution to the effective potential which is the so-called insertion contribution and given by the following two terms\cite{Dumitru:2013xna}
\be
\la{42}
U^{(2)}=g^2\left< \overline{Q_0^2}\cdot \Tr{\bf L}''(C)\right>\,
\left<{\partial\over{\partial \widehat Q_0}} \; Q^3\cdot S'''(C)\right> \,.
\ee
The first term is the renormalization of the Polyakov loop. In fact, we can calculate the expectation value of the trace of the (spatial averaged) Polyakov loop. To order $g^2$, the result is given by~\cite{Dumitru:2013xna}
\ba
<\bar{\ell}_1>&\equiv&\frac{1}{2} \Tr e^{ i \beta (C+\delta C)}\\ \nonumber
&=& \frac{1}{2} \Tr e^{i \beta C}+ i \frac{g^2}{4}\sum_{ab}\Tr(D^{aa}-D^{bb})e^{i \beta C}\frac{2+(1-\xi)}{4\pi}B_1({\cal{C}}_{ab}) \,,
\ea
where $D^{aa}$ is a diagonal matrix and the only non-vanishing element is the $a^{th}$ diagonal element which equals $1/2$. We can easily check that the traceless diagonal matrix $\delta C$ reads
\ba
\la{44}
\delta C^a=2\pi T \frac{g^2}{(4\pi)^2}(2+1-\xi)\sum_{b\neq a} B_1({\cal{C}}^{ab})\, ,
\ea
The second term in Eq.~(\ref{42}) presents the zero-momentum insertion. At two loop order, it is related to the derivatives of one-loop effective potential with respective to the constant background field $C$. As a result, the insertion contribution $U^{(2)}$ can be expressed as
\be
\la{45}
U^{(2)}=\frac{1}{\beta V}\sum_a \frac{\partial (\textrm{ln} Z^{(1)}_b+\textrm{ln} Z^{(1)}_f)}{\partial C^a} \delta C^a\, .
\ee
Notice that in the above discussions, we consider the special case where $N=2$. However, as discussed in Refs.~\cite{KorthalsAltes:1993ca,Dumitru:2013xna}, Eqs.~(\ref{44}) and (\ref{45}) are hold for any $N$.

Since the one-loop results are already known, we are able to get the insertion contribution $U_b^{(2)}$ by calculating the derivatives as shown in Eq.~(\ref{45}), the corresponding result reads
\ba
\la{46}
U_b^{(2)}=\frac{g^2T^4}{3}(2+1-\xi)\sum_{abc}B_3({\cal{C}}^{ba})B_1({\cal{C}}^{ac})\, ,
\ea
Similarly, we can get the insertion contribution for fermions
\ba
\la{47}
U_f^{(2)}&=&N_f \frac{g^2T}{2\pi}(2+1-\xi)\sum_{ab}B_1({\cal{C}}^{ab})\sumint \frac{d^4P}{(2\pi)^4}\frac{P_0^{a}}{(P^{a})^2+m^2}\nonumber\\
&\xrightarrow{m=0}& N_f\frac{g^2T^4}{3}(2+1-\xi)\sum_{ab}B_3\bigg(\frac{C^{a}-i \mu}{2\pi T}+\frac{1}{2} \bigg)B_1({\cal{C}}^{ab})\, ,
\ea

Using Eqs.~(\ref{342}) and (\ref{343}), it is clear to see that the gauge-dependent parts in the two-loop free energy are totally cancelled by the insertion contributions. The remaining terms in Eqs.~(\ref{46}) and (\ref{47}) are gauge-independent. The cancellation of the gauge-dependence in the two-loop effective potential in massless QCD has been observed in Ref.~\cite{KorthalsAltes:1993ca}. Here, we generalize this conclusion to the massive case. Notice that such a cancellation is not a trivial generalization of that in the massless limit. As we can see from Eq.~(\ref{47}), the insertion contribution from massive fermions can be simply obtained from the massless result by adding a mass term in the fermionic sum-integral, namely replacing $(P^{a})^2$ in the denominator with $(P^{a})^2+m^2$. However, the same is not true for the two-loop free energy due to the presence of the first line in Eq~(\ref{336}). The cancellation of the gauge dependence in massive QCD is guaranteed by the unexpected fact that the gauge-dependent part in the two-loop free energy, {\em i.e.}, Eq.~(\ref{343}), does satisfy the above mentioned replacement $(P^{a})^2 \rightarrow (P^{a})^2+m^2$ when going from zero to finite quark mass.

Now we are ready to write down the final result for the two-loop QCD effective potential by combining the free energy and insertion contributions. For the pure gauge part, according to Eqs.~(\ref{326}),(\ref{342}) and (\ref{46}),we have
\ba
\Gamma^{(2)}_b &=&g^2\sum_{abc}\bigg[\hat{B}_2({\cal{C}}^{ab})\hat{B}_2({\cal{C}}^{bc})+4 \hat{B}_3({\cal{C}}^{ab}) \hat{B}_1( {\cal{C}}^{ac})\bigg]-\frac{g^2NT^4}{144}\nonumber \\
&=&\frac{5g^2N^3T^4}{8\pi^4}\sum_{k=1}^{\infty}\frac{1}{k^4}\ell_k \ell_k^{*}-\frac{g^2NT^4}{144}\, .
\ea
Here, $\hat{B}_n(x)$ is defined in Eq.~(\ref{aa4}). In the second line of the above equation, when express the result in terms of the Polyakov loops, we have used the simple relation between the one- and two-loop effective potential in pure gauge theories~\cite{Guo:2014zra}. For fermionic contributions, in general, the two-loop effective potential is a sum of Eqs.~(\ref{3312}),(\ref{343}) and (\ref{47}). In the massless limit, analytical expressions are given by
\ba\la{49}
{\rm Re} \, \Gamma_f^{(2)}\big|_{m=0} &=& g^2N_f\sum_{d b}\bigg\{\bigg(\hat{B}_2({\cal{C}}^b)\hat{B}_2({\cal{C}}^d)-2\hat{B}_2({\cal{C}}^b)\hat{B}_2({\cal{C}}^{db})\bigg)\bigg(1-\frac{1}{N}\delta_{bd}\bigg)
\nonumber \\
&-&4\hat{B}_1({\cal{C}}^{db}) \hat{B}_3({\cal{C}}^d) - \bigg(\frac{\mu}{2\pi}\bigg)^2\bigg[32\pi^2 \hat{B}_1({\cal{C}}^d)\hat{B}_1({\cal{C}}^{db})\nonumber \\
&+&\bigg(\hat{B}_2({\cal{C}}^b)-\hat{B}_2({\cal{C}}^{db})+16\pi^2 \hat{B}_1({\cal{C}}^d)\hat{B}_1({\cal{C}}^b)\bigg)\bigg(1-\frac{1}{N}\delta_{bd}\bigg)\bigg]\bigg\}\nonumber \\
&+&\frac{g^2N_f}{4}\bigg(\frac{\mu}{2\pi}\bigg)^4(N^2-1)\, ,
\ea
and
\ba\la{410}
{\rm Im} \, \Gamma_f^{(2)}\big|_{m=0} &=& 4 g^2N_f \mu \sum_{d b}\bigg[\hat{B}_1({\cal{C}}^b)\bigg(\hat{B}_2({\cal{C}}^d)-\hat{B}_2({\cal{C}}^{db})\bigg)\bigg(1-\frac{1}{N}\delta_{bd}\bigg)\nonumber \\
&+&2\hat{B}_1({\cal{C}}^{db})\hat{B}_2({\cal{C}}^d)\bigg]-\frac{g^2N_f}{2\pi^2} \mu^3 \sum_b \hat{B}_1({\cal{C}}^b) (N-\frac{1}{N})\, ,
\ea

The above results can be also rewritten in terms of the Polyakov loops. However, we defer it until the relation between the one- and two-loop fermionic effective potential is obtained which allows a significant simplification. The corresponding results will be shown in section~\ref{rela}.

It is interesting to point out that in the massless limit, the one- and two-loop effective potential with non-zero quark chemical potential can be simply obtained from those at $\mu=0$ by replacing the argument ${\cal{C}}^a$ in the Bernoulli polynomials with ${\cal{C}}^a-i \frac{ \beta \mu}{2\pi}$. The definition of Bernoulli polynomials with complex arguments can be found in Appendix~\ref{appendixa}. This was first observed in Ref.~\cite{KorthalsAltes:1999cp} where explicit calculations were carried out only for the one-loop case.

Finally, we mention that a previous work\cite{Reinosa:2015gxn,Maelger:2017amh} computed the background field potential at two-loop order. In their approach, the background effective potential corresponds to the effective action evaluated in the presence of a source coupled to the background field. They used Landau-DeWitt gauge and introduced a mass term for gluon field in the action. 
The potential discussed in Refs.~\cite{Reinosa:2015gxn,Maelger:2017amh} differs from our constrained effective potential by definition, so there is no reason to directly compare the two potentials. However, formally their results at vanishing gluon mass are the same as our free energies with fixed gauge parameter $\xi=0$. In fact, the true comparison needs to be done at the level of the Polyakov loop potential, the corresponding results in Refs.~\cite{Reinosa:2015gxn,Maelger:2017amh} obtained at the next to leading order actually coincide with our constrained effective potential. This can be seen by using the  
following replacement $\tilde{C}^a \rightarrow C^a-2\pi T \frac{g^2}{(4\pi)^2}(2+1-\xi)\sum_b B_1({\cal{C}}^{ab})$\footnote{In a private communication with Urko Reinosa. Here, $\tilde{C}^a$ refers to the background field that appears in the background field potential in Refs.~\cite{Reinosa:2015gxn,Maelger:2017amh}.}. Such a replacement was first used for $SU(N)$ gauge theories\cite{Bhattacharya:1992qb,Belyaev:1991gh} in order to eliminate the $\xi$-dependence of the potential. 
In this work, we would like to show explicitly how the gauge dependence in the free energies is cancelled by that in the insertion contribution, the gauge fixing parameter $\xi$ is kept to be arbitrary in our calculations. In addition, the constrained effective potential ensures a simple proportionality between the one- and two-loop contributions for pure gauge theories. This has been discussed in Refs.~\cite{Guo:2014zra,Dumitru:2013xna}. With the obtained results, we will continue to study the similar question for fermions in next section.

\section{Relation between the one- and two-loop QCD effective potential} \la{rela}

As already known for the pure gauge part, there exists a very simple relation between the one- and two-loop constrained effective potential. Namely, two-loop correction is proportional to the one-loop result, independent on the eigenvalues of the Polyakov loop. The relation is given by
\be
\la{51}
\frac{\Gamma^{(2)}_b}{\Gamma^{(1)}_b}=-\frac{5g^2C_2(A)}{16\pi^2}\,,
\ee
where $C_2(A)$ is quadratic Casimir invariant in the adjoint representation. Remember that the insertion contribution only appears at two-loop or higher order, the effective potential at one-loop is identical to one-loop free energy. 

There is nothing in the way we perform the computation that suggests such simplicity. This proportionality was first found in a special case where the background fields lie along the edges of the Weyl chamber\cite{Bhattacharya:1992qb}. In Ref.\cite{Dumitru:2013xna}, the validity of the above relation inside the Weyl chamber was found. In fact, Eq.~(\ref{51}) is a very general result for classic groups including $G(2)$. Recently, an analytical proof of this simple relation between one- and two-loop bosonic effective potential has been completed in Ref.~\cite{Guo:2014zra}.

\begin{figure}[htbp]
\centering
\includegraphics[width=0.5\textwidth]{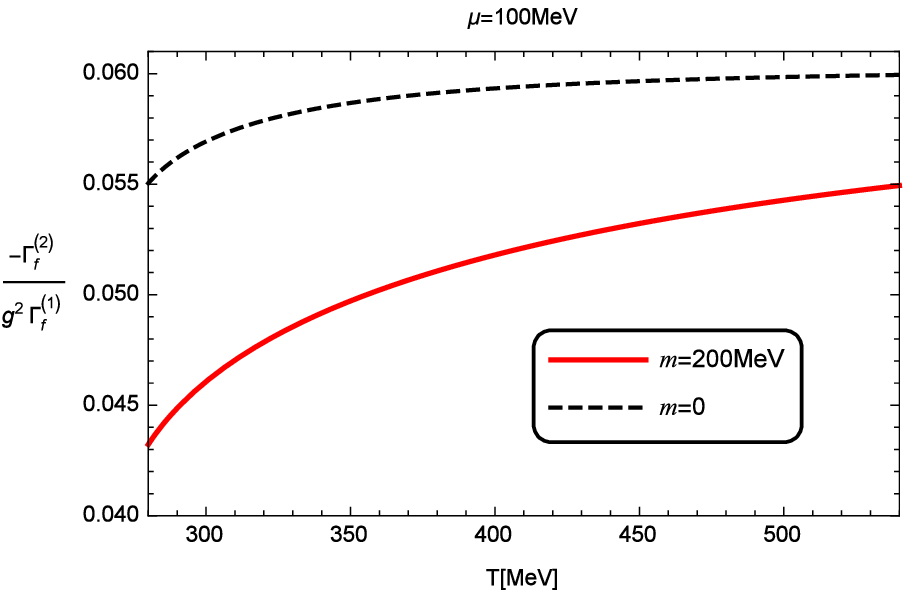}
\vspace*{-.01cm}
\caption{\small{The ratio between one- and two-loop fermionic effective potential as a function of temperature $T$. Two different quark masses $m=0$ and $m=200 {\rm MeV}$ are considered. In this plot, $N_f=1$ and the quark chemical potential $\mu=100 {\rm MeV}$. The temperature dependence of the background field is determined based on matrix models.}
\label{fig5}}
\end{figure}

It is also interesting to know if such a simple relation could hold for fermionic contributions. Therefore, we numerically evaluate the ratio $\Gamma^{(2)}_f/(g^2 \Gamma^{(1)}_f)$ which is shown in Fig.~\ref{fig5}. It is easy to see the above conclusion for the pure gauge theories doesn't work any more since the ratio becomes temperature dependent. In Fig.~\ref{fig5}, as an example, the temperature dependence of the background field is determined based on matrix models\cite{Dumitru:2012fw}. Therefore, the straight line {\em ansatz} is satisfied and $\Gamma_f$ is real even we take into account the quark chemical potential. We don't consider how the quark contributions could modify the background field which is beyond the scope of the current paper.

In Ref.~\cite{Guo:2014zra}, to prove Eq.~(\ref{51}), the basic idea is to rewrite the two-loop result in terms of $B_4(x)$ based on several relations among the periodic Bernoulli polynomials $B_n(x)$. It enables us to see the proportionality directly. We find that it is also possible to follow the same procedure for the two-loop fermionic effective potential in massless limit where analytical results have been obtained. The outcome is certainly a significant simplification of the corresponding results given in Eqs.~(\ref{49}) and (\ref{410}). In addition, for contributions that proportional to $(\frac{\mu}{T})^n$ with $n=1,2,3,4$, there also exists a similar relation for massless fermions as given by Eq.~(\ref{51}).

\subsection{The two-loop effective potential for massless quarks at $\mu=0$}

The effective potential in this limit can be easily read off from Eqs.~(\ref{49}) and (\ref{410}). For later use, we divide the sums over the color indices $b$ and $d$ into two parts: one corresponds to $b=d$ and the other is $\sum_{b\neq d}$. Then we formally express this effective potential as\footnote{In this subsection, $\Gamma^{(2)}_f$ corresponds to the effective potential at vanishing quark mass and chemical potential. To avoid the complication of the notations, we don't indicate $m=0$ and $\mu=0$ in the equations.}
\ba
\Gamma^{(2)}_f= \Gamma^{(2)}_f|_\textrm{\small{I}}+ \Gamma^{(2)}_f|_{\textrm{\small{II}}} \, ,
\ea
where
\ba
\la{p1}
\Gamma^{(2)}_f|_\textrm{\small{I}}&=& g^2N_f  \sum_b \bigg[ \hat{B}_2^2({\cal {C}}^b)-2\hat{B}_2({\cal {C}}^b)\hat{B}_2(0)   \bigg]\bigg(1-\frac{1}{N}\bigg)\, ,\\
\la{p2}
\Gamma^{(2)}_f|_{\textrm{\small{II}}}&=&g^2N_f  \sum_{b\neq d}\bigg[\hat{B}_2({\cal {C}}^b) \hat{B}_2({\cal {C}}^d)-2\hat{B}_2({\cal {C}}^b)\hat{B}_2({\cal {C}}^{db})   \bigg]\nonumber\\
&-&4g^2N_f\sum_{b\neq d}\hat{B}_3({\cal {C}}^d)\hat{B}_1({\cal {C}}^{db})     \, .
\ea

Due to the periodicity of the Bernoulli polynomials, we can assume the background fields $\pi T> C^d \ge C^b \ge -\pi T$ without losing any generality. As a result, ${\cal {C}}^b$, ${\cal {C}}^d$ and ${\cal {C}}^{db}$ are all in the interval $[0,1)$. Therefore, the corresponding expressions for $\hat{B}_n(x)$ under this assumption can be found in Eq.~(\ref{aa3}).

For $\Gamma^{(2)}_f|_\textrm{\small{I}}$, there is only one argument ${\cal {C}}^b$ in the Bernoulli polynomials, we can easily show the following relation between $\hat{B}_2(x)$ and $\hat{B}_4(x)$,
\ba
\la{id1}
\bigg[\hat{B}_2({\cal {C}}^b)-\hat{B}_2(0)\bigg]^2=\frac{3}{8\pi^2}\hat{B}_4({\cal {C}}^b)\, .
\ea
It directly leads to the simplified $\Gamma_f^{(2)}|_\textrm{\small{I}}$ as
\ba
\la{simp1}
\Gamma^{(2)}_f|_{\textrm{\small{I}}}&=& g^2N_f \sum_b\bigg[\frac{3}{8\pi^2}\hat{B}_4({\cal {C}}^b)-\frac{T^4}{144}\bigg](1-\frac{1}{N})\nonumber\\&=&g^2N_f \bigg[\sum_b \frac{N-1}{N}\frac{3}{8\pi^2}\hat{B}_4({\cal {C}}^b)-(N-1)\frac{T^4}{144}\bigg].
\ea

For $\Gamma^{(2)}_f|_\textrm{\small{II}}$, however, there are three different arguments appear in the Bernoulli polynomials. In order to re-express Eq.~(\ref{p2}) in terms of $B_4(x)$, it is very important to use the following non-trivial identity,
\ba
\la{id2}
& &\bigg[\hat{B}_2({\cal {C}}^b) \hat{B}_2({\cal {C}}^d)-2\hat{B}_2({\cal {C}}^b)\hat{B}_2({\cal {C}}^{db})-4\hat{B}_3({\cal {C}}^d)\hat{B}_1({\cal {C}}^{db})\bigg]\nonumber\\
&+&\bigg[\hat{B}_2({\cal {C}}^d) \hat{B}_2({\cal {C}}^b)-2\hat{B}_2({\cal {C}}^d)\hat{B}_2^{'}({\cal {C}}^{bd})-4\hat{B}_3({\cal {C}}^b)\hat{B}_1^{'}({\cal {C}}^{bd})\bigg]   \nonumber\\
&=&\frac{3}{8\pi^2}\bigg[\hat{B}_4({\cal {C}}^b)+\hat{B}_4({\cal {C}}^d) \bigg]-\frac{1}{8\pi^2}\hat{B}_4({\cal {C}}^{db})-\frac{T^4}{72}\, .
\ea
The above identity holds for a given pair of $(b,d)$ and the second line in the above equation has the same structure as the first line with the interchange of $b$ and $d$. Notice that we have already assume that $C^d \ge C^b$. For ${\cal {C}}^d > {\cal {C}}^b$, the argument ${\cal {C}}^{bd}$ is negative and greater than $-1$. As a result, the definition for the Bernoulli polynomials has to be modified which is indicated by a prime and we have
\ba
& & \hat{B}_2^{'}(x)=\frac{1}{2}T^2 (x^2+x+\frac{1}{6}),\nonumber\\
& & \hat{B}_1^{'}(x)=-\frac{T}{4\pi}(x+\frac{1}{2}).
\ea
In addition, we should point out the above identity is ill-defined when ${\cal {C}}^b={\cal {C}}^d$ because $B_1(x)$ has discontinuity at $x=0$. Following the analysis given in Ref.~\cite{Guo:2014zra}, due to the discontinuity, ${\cal {C}}^b={\cal {C}}^d$ should be understood as ${\cal {C}}^b={\cal {C}}^d + \epsilon$. Here, $\epsilon$ is an infinitely small (positive) number. As a result, in this special case, we have
\be
\hat{B}_3({\cal {C}}^d)\hat{B}_1({\cal {C}}^{db})+\hat{B}_3({\cal {C}}^b)\hat{B}_1^{'}({\cal {C}}^{bd})=\hat{B}_3({\cal {C}}^b)(\hat{B}_1(0^-)+\hat{B}_1(0^+))=\hat{B}_3({\cal {C}}^b)(\frac{1}{2}-\frac{1}{2})
=0\,.
\ee
As a result, Eq.~(\ref{id2}) is reduced to Eq.~(\ref{id1}) where only one argument ${\cal {C}}^b$ appears.

With the help of Eq.~(\ref{id2}), $\Gamma^{(2)}_f|_\textrm{\small{II}}$ can be simplified as
\ba
\la{simp2}
\Gamma^{(2)}_f|_\textrm{\small{II}}
 &=&g^2N_f \sum_{b>d}\bigg\{\frac{3}{8\pi^2}\bigg[\hat{B}_4({\cal {C}}^b)+\hat{B}_4({\cal {C}}^d)   \bigg]-\frac{1}{8\pi^2}\hat{B}_4({\cal {C}}^{db})-\frac{T^4}{72}\bigg\}\nonumber\\
 &=&g^2N_f\bigg[\sum_b \frac{3}{8\pi^2}\hat{B}_4({\cal {C}}^b)(N-1)-\frac{1}{8\pi^2}\sum_{b>d}\hat{B}_4({\cal {C}}^{db})-\frac{T^4}{144}(N^2-N)\bigg].
\ea

Summing up Eqs.~(\ref{simp1}) and ~(\ref{simp2}), the effective potential $\Gamma^{(2)}_f$ takes the following form which contains only Bernoulli polynomial $B_4(x)$,
\ba\label{simgf2}
\Gamma^{(2)}_f&=&\frac{3g^2N_f}{8\pi^2}\sum_b\frac{N^2-1}{N}\hat{B}_4({\cal {C}}^b)-\frac{g^2N_f}{8\pi^2}\sum_{b>d}\hat{B}_4({\cal {C}}^{db})-\frac{g^2N_fT^4}{144}(N^2-1)
\nonumber\\
&=& g^2N_fT^4\bigg[\sum_b\frac{N^2-1}{4N}B_4({\cal {C}}^b)-\frac{1}{24}\sum_{bd}B_4({\cal {C}}^{db})-\frac{1}{720}\bigg]\, .
\ea

Finally, using the one-loop result for the fermionic effective potential, we obtain
\ba
\frac{\Gamma^{(2)}_f}{\Gamma^{(1)}_f}= -\frac{3g^2}{8\pi^2}C_F(N)+\frac{g^2}{32\pi^2}\frac{\sum_{bd} B_4({\cal {C}}^{bd})+\frac{1}{30}}{\sum_b B_4({\cal {C}}^b)}\,,
\ea
where $C_F(N)=\frac{N^2-1}{2N}$. As we can see a new term $B_4({\cal {C}}^{bd})$ appears and it has no simple relation to $B_4({\cal {C}}^b)$ that appears in the one-loop effective potential. Therefore, even at vanishing quark mass, we can not expect a simple proportionality between $\Gamma^{(2)}_f$ and $\Gamma^{(1)}_f$. 

\subsection{The two-loop effective potential for massless quarks at $\mu\neq 0$}

For massless quarks, since the analytical result of $\Gamma_f$ has been obtained for finite quark chemical potential, it is also interesting to discuss the above relation for the $\mu$-dependent terms. These terms become important when the baryon density becomes larger than the temperatures. The leading contribution is very simply which is proportional to $(\mu/T)^4$ and independent on the background field, it is straightforward to show that\footnote{$\Gamma^{(2)}_f$ and $\Gamma^{(1)}_f$ in this subsection refer to terms proportional to either $(\mu/T)^4$ or $(\mu/T)^2$ in the two- and one-loop fermionic effective potential, respectively. The actual meaning becomes clear according to the text.}
\ba
\la{refer}
\frac{\Gamma^{(2)}_f}{\Gamma^{(1)}_f}= -\frac{3g^2}{8\pi^2}C_F(N)\,.
\ea

Next we consider the contributions that proportional to $(\mu/T)^2$. According to the analytical expression for the one-loop effective potential in Eq.~(\ref{3117}), we need to re-express the corresponding two-loop result Eq.~(\ref{49}) in terms of $\hat{B}_2(x)$. The idea is very similar as that in previous subsection, we first need to divide the sums over $b$ and $d$ into the following two parts
\be
\Gamma^{(2)}_f=\Gamma^{(2)}_f|_\textrm{\small{I}}+\Gamma^{(2)}_f|_\textrm{\small{II}}\,,
\ee
where
\ba
\Gamma^{(2)}_f|_\textrm{\small{I}}&=& -\frac{g^2N_f T^2}{4\pi^2}\bigg(\frac{\mu}{T}\bigg)^2 \sum_{b}\bigg[ \hat{B}_2({\cal {C}}^b)- \hat{B}_2(0)+16\pi^2\hat{B}_1({\cal {C}}^b)\hat{B}_1({\cal {C}}^b)\bigg]\bigg(1-\frac{1}{N}\bigg)\, ,\nonumber\\
\Gamma^{(2)}_f|_\textrm{\small{II}}&=& -\frac{g^2N_f T^2}{4\pi^2}\bigg(\frac{\mu}{T}\bigg)^2\sum_{b\neq d}\bigg[\hat{B}_2({\cal {C}}^b)- \hat{B}_2({\cal {C}}^{db})+16\pi^2\hat{B}_1({\cal {C}}^d)\hat{B}_1({\cal {C}}^b) \bigg]\, \nonumber\\
&-&8g^2N_f T^2\bigg(\frac{\mu}{T}\bigg)^2\sum_{b\neq d} \hat{B}_1({\cal {C}}^d) \hat{B}_1({\cal {C}}^{db})\,.
\ea
With the help of the following two identities\footnote{Since the proof of these two identities is very similar as that of Eqs.~(\ref{id1}) and (\ref{id2}), we don't show the details.}
\be
\hat{B}_1({\cal {C}}^b)\hat{B}_1({\cal {C}}^b)=\frac{1}{8\pi^2}\bigg[\hat{B}_2({\cal {C}}^b)+\frac{T^2}{24}\bigg]\, ,
\ee
and
\ba
&&\bigg[\hat{B}_2({\cal {C}}^b)- \hat{B}_2({\cal {C}}^{db})+16\pi^2\hat{B}_1({\cal {C}}^d)\hat{B}_1({\cal {C}}^b)+ 32\pi^2 \hat{B}_1({\cal {C}}^d) \hat{B}_1({\cal {C}}^{db})\bigg] \nonumber\\
&+& \bigg[\hat{B}_2({\cal {C}}^d)- \hat{B}^\prime_2({\cal {C}}^{bd})+16\pi^2\hat{B}_1({\cal {C}}^b)\hat{B}_1({\cal {C}}^d)+ 32\pi^2 \hat{B}_1({\cal {C}}^b) \hat{B}^\prime_1({\cal {C}}^{bd})\bigg]  \nonumber\\
&=& 3\bigg[B_2({\cal {C}}^d)+B_2({\cal {C}}^b)\bigg]\,,
\ea
we arrive at
\ba
\Gamma_f^{(2)}|_\textrm{\small{I}}&=& -\frac{g^2N_f T^2}{4\pi^2}\bigg(\frac{\mu}{T}\bigg)^2 3\sum_{b} \hat{B}_2({\cal {C}}^b)\bigg(1-\frac{1}{N}\bigg)\, ,\nonumber\\
\Gamma_f^{(2)}|_\textrm{\small{II}}&=& -\frac{g^2N_f T^2}{4\pi^2}\bigg(\frac{\mu}{T}\bigg)^2 3\sum_{b > d}\bigg[\hat{B}_2({\cal {C}}^b)+\hat{B}_2({\cal {C}}^d)\bigg]\, \nonumber\\
&=& -\frac{g^2N_f T^2}{4\pi^2}\bigg(\frac{\mu}{T}\bigg)^2 3\sum_{b}\hat{B}_2({\cal {C}}^b)(N-1)\,,
\ea
and
\be
\Gamma_f^{(2)}=\Gamma_f^{(2)}|_\textrm{\small{I}}+\Gamma_f^{(2)}|_\textrm{\small{II}}=-\frac{3g^2N_f T^2}{2\pi^2}\bigg(\frac{\mu}{T}\bigg)^2C_F(N)\sum_{b}\hat{B}_2({\cal {C}}^b)\, .
\ee
Using the one-loop result, we find the exactly same relation as given by Eq.~(\ref{refer}).

As we know, when considering the background field at finite quark chemical potential, the effective potential is in general complex. In the massless limit, the imaginary parts contain terms proportional to
$(\mu/T)^3$ and $\mu/T$. Although these contributions are not physical, we can still analytically check if Eq.~(\ref{refer}) also holds for them. Surprisingly we do find the same proportionality exists. For terms proportional to $(\mu/T)^3$, both one- and two-loop contributions are simply proportional to $B_1(x)$, therefore, it is trivial to show the proportionality. On the other hand, for terms proportional to $\mu/T$, one can use the same approach as above to get the corresponding relation. Here we only list the following key identities for readers who want to go through the details.
\be
\hat{B}_1({\cal {C}}^b)\bigg[\hat{B}_2({\cal {C}}^b)-\hat{B}_2(0)\bigg]=-\frac{3}{16\pi^2}\hat{B}_3({\cal {C}}^b)\, ,
\ee
\ba
&&\bigg[\hat{B}_1({\cal {C}}^b)\hat{B}_2({\cal {C}}^d)- \hat{B}_1({\cal {C}}^b)\hat{B}_2({\cal {C}}^{db})+2\hat{B}_2({\cal {C}}^d)\hat{B}_1({\cal {C}}^{db})\bigg] \nonumber\\
&+& \bigg[\hat{B}_1({\cal {C}}^d)\hat{B}_2({\cal {C}}^b)- \hat{B}_1({\cal {C}}^d)\hat{B}^\prime_2({\cal {C}}^{bd})+2\hat{B}_2({\cal {C}}^b)\hat{B}^\prime_1({\cal {C}}^{bd})\bigg]  \nonumber\\
&=&-\frac{3}{16\pi^2}\bigg[B_3({\cal {C}}^d)+B_3({\cal {C}}^b)\bigg]\,.
\ea

Given the above relations between the one- and two-loop fermionic effective potential, we can also express Eqs.~(\ref{49}) and (\ref{410}) in terms of the Polyakov loops. The imaginary part as well as the $\mu$-dependent terms in the real part of the two-loop effective potential can be simply obtained from Eq.~(\ref{3117add}). For the $\mu$-independent terms in the real part, we can use the simplified expression in Eq.~(\ref{simgf2}) and change the Bernoulli polynomials into the Polyakov loops. The final results are the following,
\ba
{\rm Re} \, \Gamma_f^{(2)}\big|_{m=0} &=& \frac{g^2 N_f  T^4}{64\pi^4}\bigg\{(N^2-1)\bigg[\bigg(\frac{\mu}{T}\bigg)^4-24\bigg(\frac{\mu}{T}\bigg)^2\sum_{k=1}^{\infty} \frac{(-1)^k}{k^2}
{\rm Re}\,\ell_k-48\sum_{k=1}^{\infty} \frac{(-1)^k}{k^4}{\rm Re}\,\ell_k\bigg]\nonumber\\
&+&8N^2\sum_{k=1}^{\infty} \frac{1}{k^4}\ell_k\,\ell_k^*-\frac{4\pi^4}{45}\bigg\}\, ,\\
{\rm Im} \, \Gamma_f^{(2)}\big|_{m=0} &=& -\frac{g^2 N_f  T^4}{8\pi^4}(N^2-1) \bigg(\frac{\mu}{T}\bigg)\sum_{k=1}^{\infty}\bigg[\bigg(\frac{\mu}{T}\bigg)^2+\bigg(\frac{6}{k^2}\bigg)\bigg]\frac{(-1)^k}{k}{\rm Im}\,\ell_k\, .
\ea

To conclude, we find that even at vanishing quark mass, in general, there is no simple proportionality between the one- and two-loop fermionic effective potential. However, at high baryon chemical potential and low temperatures, up to ${\cal O}(\frac{\mu}{T})$, the ratio $\Gamma_f^{(2)}/\Gamma_f^{(1)}$ becomes independent on the eigenvalues of the Polyakov loop\footnote{For the fermionic effective potential, in the massless limit, the contributions up to ${\cal O}(\frac{\mu}{T})$ are actually the terms dependent on the chemical potential $\mu$. To be more clear, they refer to $\Gamma_f(\mu,m=0)-\Gamma_f(\mu=0,m=0)$.}. This is very similar as the relation holds for pure gauge theories. Instead of  $-\frac{5g^2}{16\pi^2}C_2(A)$ as given in Eq.~(\ref{51}), the ratio becomes $-\frac{3g^2}{8\pi^2}C_F(N)$ for the fermionic case.

\section{Summary and Outlook}\la{summary}

In this work, we have perturbatively computed the (constrained) effective potential of QCD up to two-loop order by considering a constant background field $A_0^{{\rm cl}}$. It extended the previous studies for pure gauge theories to the full QCD case with finite quark mass and chemical potential. For massless quarks, the relation between the one- and two-loop contributions has also been studied analytically.

We adopted the double line notations to deal with the color structures and developed some new techniques to perform the Matsubara frequency sums which simplify the calculations to some extent. Our results show that the one-loop free energy in a constant background field can be expressed with the same functional form as that at $A_0^{{\rm cl}}=0$ if we introduce the $A_0^{{\rm cl}}$ dependence into the Fermi-Dirac and Bose-Einstein distributions. However, at two-loop order, this conclusion only holds in the Feynman gauge for bosonic contributions. For fermionic free energy, the same conclusion can be drawn only in the massless limit because a new function as given by Eq.~(\ref{3314}) arises when finite quark mass is taken into account. In addition, the free energy as well as the effective potential can be evaluated in closed form for massless quarks which has been obtained in terms of the periodic Bernoulli polynomials or Polyakov loops.

The constrained action in which one keeps the value of the Polyakov loop to be fixed in the path integral leads to an insertion contribution to the effective potential at two-loop order. With our explicit results, we show that the two-loop free energy becomes gauge-dependent after including a background field. On the other hand, the gauge dependence in the insertion contribution exactly cancels that in the free energy and the constrained effective potential is a gauge-independent quantity as expected. After considering the quark mass, although some new term appears in the free energy, the way how the gauge dependence is cancelled does not change as compared to the massless case. Furthermore, we found that unlike the pure gauge theories, in general, there is no simple proportionality between the one- and two-loop fermionic effective potential. On the other hand, when neglecting the quark mass, the $\mu$-dependent terms in the two-loop correction are proportional to those in the one-loop result. The proportional coefficient is independent on the eigenvalues of the Polyakov loop, but different from that for pure gauge theories.

It is also very interesting to consider the adjoint Weyl fermions instead of fundamental Dirac fermions, then one has the ${\cal N}=1$ Supersymmetric Yang-Mills theory. As shown in Ref.~\cite{Poppitz:2012sw}, when the periodic boundary conditions are imposed for fermions, the potential induced by (massless) fermions at vanishing chemical potential must cancel the gluonic potential. As a result, one can expect the two-loop result for adjoint fermions must be proportional to corresponding one-loop result, just like glue sector.

Our results can be used to improve the matrix models for QCD phase transition which typically employ the one-loop effective potential. Besides corrections to the perturbative contributions in these models, it is also expected to suggest some possible fermionic terms that contribute non-perturbatively to the matrix models. This is achieved when one includes a mass term in the dispersion relation of gauge bosons and these non-perturbative contributions from fermions have not been discussed in present models. In addition, our finding on the relation between the one- and two-loop fermionic effective potential may provide some useful information for studying the QCD phase diagram at high baryon densities. One example is the equation of state for compact stars. Finally, we would like to mention that it is also interesting to go beyond two-loop order and there are already some partial results in a very recent work~\cite{Nishimura:2018wla}. The calculation involves some new technical problems since higher order contributions may contain infrared divergences which don't show up in present work. The resummation of these divergence leads to higher order corrections at $g^3$. Furthermore, some theoretically important questions could be answered with the corresponding results. For instance, the way how the gauge dependence is cancelled between the free energy and insertion contribution may become complicated and it should be checked explicitly which ensures the consistency of the calculation. In addition, one may also wondering if the above mentioned proportionality is a general conclusion that may also exist in higher order contributions. We postpone these studies in the future work.

\subsection*{Acknowledgments}
\label{sec:acknowledgments}

 This work is supported by the NSFC of China under Project No. 11665008, by Natural Science Foundation of Guangxi Province of China under Project Nos. 2016GXNSFFA380014, 2018GXNSFAA138163 and by the ¡°Hundred Talents Plan¡± of Guangxi Province of China.

\appendix

\section{Periodic Bernoulli polynomials}
\la{appendixa}

We define the periodic Bernoulli polynomials,
\ba
B_{2l}(x)=\sum_{n=1}^\infty (-1)^{l-1} \frac{2(2l)!}{(2\pi n)^{2l}}\textrm{cos}(2\pi x n)\, ,
\ea
which satisfy
\ba
2l B_{2l-1}(x)=B_{2l}^{'}(x)\, .
\ea

It is easy to show that the above defined Bernoulli polynomials are periodic functions of $x$, with period $1$. For $0\le x \le 1$, the explicit forms reads
\ba\la{aa3}
& &B_4(x)=x^2(1-x)^2-\frac{1}{30}\, ,\nonumber\\
& &B_3(x)=x^3-{\frac{3}{2}}x^2+\frac{1}{2}x\, ,\nonumber\\
& &B_2(x)=x^2-x+\frac{1}{6}\, ,\nonumber\\
& &B_1(x)=x-\frac{1}{2}\, .
\ea
For arbitrary values of $x$, the argument of the above Bernoulli polynomials should be understood as $x-[x]$ with $[x]$ the largest integer less
than 
$x$, which is nothing but the modulo function. We should point out that $B_1(x)$ has discontinuities at integer $x$. For example, the value of $B_1(x)$ at $x = 1$ depends on the way how $x$ approaches zero, from above or from below. We have $B_1(1^+)=-\frac{1}{2}$ and $B_1(1^-)=\frac{1}{2}$.

In addition, we also define $\hat{B}_n(x)$ with $n=1,2,3,4$ which have a simple relation to $B_n(x)$,
\ba\la{aa4}
& &\widehat B_4(x)=\frac{2}{3}\pi^2T^4 \bigg(B_4(x)+\frac{1}{30}\bigg)\,,\nonumber\\
& &\widehat B_3(x)=\frac{2}{3}\pi T^3 B_3(x)\,,\nonumber\\
& &\widehat B_2(x)=\frac{1}{2}T^2 B_2(x)\,,\nonumber\\
& &\widehat B_1(x)=-\frac{T}{4\pi} B_1(x)\,.
\ea

In our calculation, we will also encounter the Bernoulli polynomials with complex argument $B_n(x+i y)$. It is defined as
\ba
\la{bercpl}
B_n(x+i y) = -\frac{n!}{(2\pi i)^n} \sum_{m=0}^n \frac{(- 2\pi y)^{n-m}}{(n-m)!}\mathop{{\sum}'}_{k=-\infty}^{\infty} k^{-m} e^{2\pi i k x}\, ,
\ea
where $\sum^\prime$ indicates the $k=0$ term in the summation should be dropped. 
In fact, from Eq.~(\ref{bercpl}), we can easily check that Eq.~(\ref{aa3}) also holds for complex argument $z=x+i y$ with $0\le x \le1$.

\section{Bosonic and fermionic sum-integrals}\la{appendixB}

In this appendix, we derive the sum-integrals that can be used to compute the two-loop effective potential in a constant background field. We start with the bosonic sum-integral which can be expressed as
\ba
\la{ab1}
\sumint\frac{d^4Q}{(2\pi)^4}\frac{1}{(Q^{ab})^2}&=&T\sum_n\int_{\textbf{q}}\frac{1}{(2\pi n T+C^{ab})^2+q^2}\nonumber \\
&=&\frac{i}{4\pi}\sum_n\int_{\textbf{q}}\frac{1}{q}\bigg[\frac{1}{n+ {\cal {C}}^{ab}+ i \beta q/(2\pi)}- \frac{1}{n+ {\cal {C}}^{ab} - i \beta q/(2\pi)}\bigg]\,.\nonumber \\
\ea
The Matsubara frequency sum can be carried out by using the basic identity Eq.~(\ref{316}), which leads to
\ba
\la{ab2}
\sumint\frac{d^4Q}{(2\pi)^4}\frac{1}{(Q^{ab})^2}&=&\frac{1}{4}\int_{\textbf{q}}\frac{1}{q} \bigg[\coth\bigg(\frac{q+i C^{ab}}{2T}\bigg)-\coth\bigg(\frac{-q+i C^{ab}}{2T}\bigg)\bigg]\,,\nonumber \\
&=&\frac{1}{4\pi^2}\int q dq (N_{ab}^+(q)+N_{ab}^-(q)+1)\,.
\ea
In the above equation, we used $\coth(z)=1+2/(e^{2z}-1)=-1-2/(e^{-2z}-1)$. After integrating over $q$, the final result of the bosonic sum-integral reads
\ba
\la{ab3}
\sumint\frac{d^4Q}{(2\pi)^4}\frac{1}{(Q^{ab})^2}= \frac{T^2}{2}B_2({\cal {C}}^{ab})+\frac{1}{4\pi^2}\int q dq\, .
\ea

For fermions, the fermionic sum-integral can be derived with exactly the same procedure as above and we only list the corresponding result.
\ba
\la{ab4}
\sumint\frac{d^4P}{(2\pi)^4}\frac{1}{(P^a)^2+m^2}&=&T\sum_n\int_{\textbf{p}}\frac{1}{[(2n+1) \pi T+C^a-i\mu]^2+p^2+m^2}\nonumber \\
&=&-\frac{1}{4\pi^2}\int \frac{p^2dp}{\sqrt{p^2+m^2}}  (N_{a}^+(p)+N_{a}^-(p)-1)\, .
\ea

When $m=0$, the above equation can be expressed in terms of the periodic Bernoulli polynomials
\ba
\la{ab5}
\sumint\frac{d^4P}{(2\pi)^4}\frac{1}{(P^a)^2}&=&\frac{T^2}{2}B_2({\cal{C}}^a-i\frac{\beta\mu}{2\pi})+\frac{1}{4\pi^2}\int p dp\nonumber\\&=&\frac{T^2}{2}B_2({\cal{C}}^a)-\frac{T^2}{8\pi^2}(\frac{\mu}{T})^2-i \frac{T^2}{2\pi}\frac{\mu}{T}B_1({\cal{C}}^a)+\frac{1}{4\pi^2}\int p dp
\, .
\ea

Eqs.~(\ref{ab2}) and (\ref{ab4}) are related to the sum of the parton distribution functions, $N_{ab}(q)=N_{ab}^+(q)+N_{ab}^-(q)$ or $N_{a}(p)=N_{a}^+(p)+N_{a}^-(p)$. In fact, there are some other sum-integrals which are related to the difference of the parton distribution functions, $\bar{N}_{ab}(q)=N_{ab}^+(q)-N_{ab}^-(q)$ or $\bar{N}_{a}(p)=N_{a}^+(p)-N_{a}^-(p)$. For non-zero background field, these sum-integrals will contribute to the free energy at two-loop order which make the result gauge dependent. With the same approach, we obtain the following expressions
\be
\la{ab6}
\sumint\frac{d^4Q}{(2\pi)^4}\frac{Q_0^{ab}}{(Q^{ab})^2}=\frac{-i}{4\pi^2}\int q^2 dq (N_{ab}^+(q)-N_{ab}^-(q))=\frac{2\pi T^3}{3}B_3({\cal {C}}^{ab})\, ,
\ee
\be
\la{ab7}
\sumint\frac{d^4P}{(2\pi)^4}\frac{P_0^{a}}{(P^{a})^2+m^2}=\frac{i}{4\pi^2}\int p^2 dp (N_{a}^+(p)-N_{a}^-(p))
\xrightarrow{m=0}\frac{2\pi T^3}{3}B_3\bigg({\cal {C}}^{a}-i \frac{ \beta \mu}{2\pi}\bigg)\, ,
\ee
\be
\la{ab8}
\sumint\frac{d^4Q}{(2\pi)^4}\frac{Q_0^{ab}}{(Q^{ab})^4}=\frac{-i}{8\pi^2}\int  dq (N_{ab}^+(q)-N_{ab}^-(q))=-\frac{T}{4\pi}B_1({\cal {C}}^{ab})\, .
\ee
We should mention that Eq.~(\ref{316}) can not be directly used to perform the Matsubara frequency sum in Eq.~(\ref{ab8}). In fact, the simple way to get this equation is to take derivatives with respective to the background field on both sides of Eq.~(\ref{ab3}).

\section{Matsubara frequency sum with finite quark mass}\la{appendixC}
We want to calculate the following quantity ${\cal S}$ which is given by
\ba
\la{ac1}
{\cal S}&=& \sum_{db}\sumint \frac{d^4P}{(2\pi)^4}\frac{d^4K}{(2\pi)^4}\frac{d^4Q}{(2\pi)^4}\frac{(2\pi)^4\delta^4(P^d-K^b+Q^{bd})}
{[(K^{b})^2+m^2][(P^{d})^2+m^2](Q^{bd})^2}\nonumber \\
&=&\beta^4\sum_{db} \sum_{n_p ,n_q} \int_{{\bf p}} \int_{{\bf k}} \frac{1}{[(2n_p+1)\pi +\alpha]^2+x^2}\frac{1}{(2\pi n_q+\gamma)^2+z^2}\nonumber \\
&\times & \frac{1}{[(2n_p+1)\pi+2\pi n_q+\rho]^2+y^2}\, .
\ea
Notice that in the above equation, we sum over $n_k$ and perform the integration $\int d{{\bf q}}$ with the help of the delta function. To keep the notations compact, we also introduce the following dimensionless variables
\ba\la{ac2}
&&\alpha=\beta C^d-i \beta \mu\, ,\quad \quad \rho=\beta C^b-i \beta \mu\, ,\quad \quad \gamma=\rho-\alpha\, ,\nonumber \\
&&x=E_p/T\, ,\quad\quad\quad \quad \, y=E_k/T\, ,\quad\quad\quad \quad\,\, z=|{\bf {k}}-{\bf {p}}|/T\equiv \omega/T\, .
\ea

The first step is to carry out the Matsubara frequency sum over $n_p$.
\ba\la{ac3}
&&\sum_{n_p}\frac{1}{(2\pi n_p+\pi+\alpha)^2+x^2}\frac{1}{[2\pi (n_p+n_q)+\pi+\rho]^2+y^2}\nonumber \\
&=&-\frac{1}{16\pi^2xy}\sum_{n_p}\bigg(\frac{1}{n_p+\frac{\pi+\alpha+ix}{2\pi}}-\frac{1}{n_p+\frac{\pi+\alpha-ix}{2\pi}}\bigg)
\bigg(\frac{1}{n_p+\frac{2 \pi n_q+\pi+\rho+iy}{2\pi}}-\frac{1}{n_p+\frac{2 \pi n_q+\pi+\rho-iy}{2\pi}}\bigg)\nonumber \\
&=&\frac{i}{4xy}\bigg[\frac{N_b^+(k)-N_d^+(p)}{2\pi n_q+\gamma-i(x-y)}+\frac{N_d^-(p)-N_b^-(k)}{2\pi n_q+\gamma+i(x-y)}+\frac{N_b^-(k)+N_d^+(p)-1}{2\pi n_q+\gamma-i(x+y)}\nonumber \\
&&\quad\quad+\frac{1-N_d^-(p)-N_b^+(k)}{2\pi n_q+\gamma+i(x+y)}\bigg]\, ,
\ea
There are four terms in the second line of the above equation and the corresponding frequency sums can be simply obtained by using Eq.~(\ref{3311}).

The second step is to perform the sum over $n_q$ and there are four similar sums we need to consider
\ba
\la{ac4}{\rm I}&=&\sum_{n_q}\frac{i}{4xy}\frac{1}{(2\pi n_q+\gamma)^2+z^2}\frac{N_b^+(k)-N_d^+(p)}{2\pi n_q+\gamma-i(x-y)}\, , \\
\la{ac5}{\rm II}&=&\sum_{n_q}\frac{i}{4xy}\frac{1}{(2\pi n_q+\gamma)^2+z^2}\frac{N_d^-(p)-N_b^-(k)}{2\pi n_q+\gamma+i(x-y)}\, , \\
\la{ac6}{\rm III}&=&\sum_{n_q}\frac{i}{4xy}\frac{1}{(2\pi n_q+\gamma)^2+z^2}\frac{N_b^-(k)+N_d^+(p)-1}{2\pi n_q+\gamma-i(x+y)}\, , \\
\la{ac7}{\rm IV}&=&\sum_{n_q}\frac{i}{4xy}\frac{1}{(2\pi n_q+\gamma)^2+z^2}\frac{1-N_d^-(p)-N_b^+(k)}{2\pi n_q+\gamma+i(x+y)}\, .
\ea
In order to avoid the rather tedious calculations, it is important to point out the following facts. Although the result of Eq.~(\ref{ac6}) is not same as Eq.~(\ref{ac7}), they have identical contributions to ${\cal S}$. This is clear to see when we interchange ${\bf p}$ and ${\bf k}$ as well as the color indices $b$ and $d$. Using the above standard techniques for the Matsubara frequency sum, we have
\ba\la{ac8}
&&\sum_{n_q}\frac{1}{(2\pi n_q+\gamma)^2+z^2}\frac{1}{2\pi n_q+\gamma+i(x+y)}\nonumber\\
&=&\frac{- i}{2 z}\bigg[\frac{N_{bd}^-(\omega)+N_{bd}^+(E_p+E_k)+1}{x+y+z}+\frac{N_{bd}^+(\omega)-N_{bd}^+(E_p+E_k)}{x+y-z}\bigg]\, ,
\ea
where
\be\la{ac9}
N_{bd}^+(E_p+E_k)=\frac{N_d^-(p)N_b^+(k)}{1-N_d^-(p)-N_b^+(k)}\, .
\ee
Then, it is straightforward to compute Eq.~ (\ref{ac7}) and we get
\ba\la{ac10}
{\rm IV}&=&\frac{- 1}{8 x y z}\bigg[\bigg(\frac{N_{bd}^-(\omega)+1}{x+y+z}+\frac{N_{bd}^+(\omega)}{x+y-z}\bigg)(N_d^-(p)+N_b^+(k)-1)\nonumber\\
&+&\frac{2z}{(x+y)^2-z^2}N_d^-(p)N_b^+(k)\bigg]
\ea
By interchanging the integral variables and the color indices $b$ and $d$, a more useful expression for Eq.~ (\ref{ac7}) can be obtained.
\ba\la{ac11}
{\rm IV}&\rightarrow&\frac{- 1}{8 x y z}\bigg[\frac{(x+y)N_{bd}(\omega) N_d(p)-z\bar{N}_{bd}(\omega) \bar{N}_d(p)}{(x+y)^2-z^2}-\frac{2y N_{bd}(\omega)}{(x+y)^2-z^2}\nonumber\\
&+&z \frac{N_d^-(p)N_b^+(k)+N_d^+(p)N_b^-(k)}{(x+y)^2-z^2}+\frac{N_d(p)-1}{x+y+z}\bigg]
\ea
In the above equation, the right arrow indicates that the right side of this equation is not equal to Eq.~ (\ref{ac7}), however, it gives the same contribution to $\cal{S}$. The distribution functions $\bar{N}_{bd}(\omega)$ and  $\bar{N}_d(p)$ are defined under Eq.~(\ref{3316}).

The above procedure also applies when we compute the sum in Eq.~ (\ref{ac4}) and we have
\ba\la{ac12}
{\rm I}&=&\frac{- 1}{8 x y z}\bigg[\frac{N_{bd}^+(\omega)(N_b^+(k)-N_d^+(p))+N_b^+(k)}{x-y+z}+2z\frac{N_b^+(k)N_d^+(p)}{(x-y)^2-z^2}\nonumber\\
&+&\frac{N_{bd}^-(\omega)(N_b^+(k)-N_d^+(p))-N_d^+(p)}{x-y-z}\bigg]
\ea
Furthermore, we can read off the result for Eq.~(\ref{ac5}) from Eq.~(\ref{ac12}) by using the following changes of the dimensionless variables as defined in Eq.~(\ref{ac2})
\ba\la{ac13}
{\rm II}={\rm I}(C^b\rightarrow -C^d, C^d\rightarrow -C^b, \mu\rightarrow -\mu, {\bf p}\rightarrow {\bf k}, {\bf k} \rightarrow {\bf p})\,.
\ea
Then the total contributions from Eqs.~(\ref{ac4}) and (\ref{ac5}) to the quantity $\cal{S}$ can be written as
\ba\la{ac14}
{\rm I}+{\rm II}&\rightarrow&\frac{- 1}{4 x y z}\bigg[\frac{(y-x)N_{bd}(\omega) N_d(p)+z\bar{N}_{bd}(\omega) \bar{N}_d(p)}{(x-y)^2-z^2}\nonumber\\
&+&z \frac{N_d^-(p)N_b^-(k)+N_d^+(p)N_b^+(k)}{(x-y)^2-z^2}-\frac{N_d(p)}{x-y-z}\bigg]
\ea

We should point out that terms related to $N_{bd}(\omega) N_d(p)$ don't contribute to ${\cal S}$. This can be verified by changing ${\bf k}$ into ${\bf p+k}$ and integrating over the polar angle $d \Omega_k$. However, the terms related to $\bar{N}_{bd}(\omega) \bar{N}_d(p)$ do have a contribution to ${\cal S}$ and this contribution vanishes at zero background field by definition.

Summing up all the contributions, we finally get
\ba
\la{ac15}
{\cal S}&=& -\sum_{db} \int_{{\bf p}} \int_{{\bf k}} \frac{1}{4E_pE_k} \bigg\{\frac{N_b^+(k)N_d^-(p)+N_b^-(k)N_d^+(p)-\bar{N}_{bd}(\omega) \bar{N}_d(p)}{(E_k+E_p)^2-\omega^2}\nonumber \\
&+ & \frac{N_b^+(k)N_d^+(p)+N_b^-(k)N_d^-(p)+\bar{N}_{bd}(\omega) \bar{N}_d(p)}{(E_k-E_p)^2-\omega^2}-\frac{2 E_k N_{bd}(\omega)}{\omega [(E_k+E_p)^2-\omega^2]}\nonumber \\
&+&\frac{N_d(p)}{\omega}\bigg(\frac{1}{E_k+E_p+\omega}-\frac{1}{E_p-E_k-\omega}\bigg)-\frac{1}{\omega}\frac{1}{E_k+E_p+\omega}\bigg\}\, .
\ea

\bibliographystyle{JHEP}

\providecommand{\href}[2]{#2}\begingroup\raggedright

\endgroup

\end{document}